\documentclass[twocolumn,showpacs,amsmath,amssymb,pre,aps]{revtex4}   
\usepackage{graphicx}
\usepackage{dcolumn}
\usepackage{bm}

\begin{document}
\title{
Critical currents in superconductors with quasiperiodic pinning arrays: 
One-dimensional chains and two-dimensional Penrose lattices 
}
\draft

\author{Vyacheslav Misko$^{1, 2}$, Sergey Savel'ev$^1$, and Franco Nori$^{1,2}$}
\affiliation{$^1$ Frontier Research System, The Institute of Physical and Chemical 
Research (RIKEN), Wako-shi, Saitama, 351-0198, Japan} 
\affiliation{$^2$ Center for Theoretical Physics,
Center for the Study of Complex Systems, Department of Physics,
University of Michigan, Ann Arbor, MI 48109-1040, USA}

\date{\today}

\begin{abstract}
We study the critical depinning current $J_{c}$, as a function of the applied 
magnetic flux $\Phi$, for quasiperiodic (QP) pinning arrays, including one-dimensional (1D) chains 
and two-dimensional (2D) arrays of pinning centers placed on the nodes of a five-fold 
Penrose lattice.
In 1D QP chains of pinning sites,
the peaks in $J_{c}(\Phi)$ are shown to be determined by a sequence of harmonics 
of long and short periods of the chain. 
This sequence includes as a subset the sequence of successive Fibonacci numbers. 
We also analyze the evolution of $J_{c}(\Phi)$ 
while a continuous transition occurs from a periodic lattice of pinning centers to a QP one; 
the continuous transition is achieved by varying the ratio $\gamma = a_{S}/a_{L}$ of lengths 
of the short $a_{S}$ and the long $a_{L}$ segments, starting from $\gamma = 1$ for a periodic 
sequence. 
We find that the peaks related to the Fibonacci sequence are most pronounced when 
$\gamma$ 
is equal to the ``golden mean''. 
The critical current $J_{c}(\Phi)$ in QP lattice has a remarkable self-similarity.
This effect is demonstrated both in real space and in reciprocal $k$-space. 
In 2D QP pinning arrays (e.g., Penrose lattices), the pinning of vortices is 
related to matching conditions between the vortex lattice and the QP 
lattice of pinning centers. 
Although more subtle to analyze than in 1D pinning chains, the structure in $J_{c}(\Phi)$ 
is determined by the presence of two different kinds of elements forming the 2D QP lattice. 
Indeed, we predict analytically and numerically 
the main features of $J_{c}(\Phi)$ for Penrose lattices. 
Comparing the $J_{c}$'s for QP (Penrose), periodic (triangular) and random 
arrays of pinning sites, 
we have found 
that the QP lattice 
provides an unusually broad critical current $J_{c}(\Phi)$, 
that could be useful for practical applications 
demanding high 
$J_{c}$'s 
over a wide range of fields. 
\end{abstract}
%
\maketitle


\section{Introduction}

Recent progress in the fabrication of nanostructures 
has provided a wide variety of 
well-controlled vortex-confinement topologies, 
including different types of regular pinning arrays. 
A main fundamental question 
in this field is how to drastically increase vortex pinning, 
and thus the critical current 
$J_{c}$, 
using artificially-produced periodic arrays of pinning sites (APS). 
These periodic APS have been extensively used for studying 
vortex pinning and vortex dynamics.
In particular, 
enhanced 
$J_{c}$ 
and 
commensurability effects have been demonstrated in superconducting thin films with 
square and triangular 
arrays of sub-$\mu$m holes (i.e., antidots) 
\cite{vvmdotprl,vvmdotprb,fnsc2003,rwdot,ulm01,ulm02}. 
Moreover, 
blind antidots 
(i.e., 
holes which partially perforate the film to a certain depth) 
\cite{vvmbdot}, 
or 
pinning arrays with field-dependent pinning strength 
\cite{vvmfddot}, 
provide more flexibility for controlling 
properties such as pinning strength, anisotropy, etc. 
The increase and, more generally, control 
of the critical current 
$J_{c}$ 
in superconductors 
by its patterning (perforation)
can be of practical importance for applications
in micro- and nanoelectronic devices. 

A peak in the critical current 
$J_{c}(\Phi)$, 
for a given value of the magnetic flux, 
say 
$\Phi_{1}$, 
can be engineered using a superconducting sample 
with a periodic APS 
with a matching field 
$H_{1}=\Phi_{1}/A$
(where $A$ is the area of the pinning cell), 
corresponding to one trapped vortex per pinning site. 
However, this peak in 
$J_{c}(\Phi)$, 
while useful to obtain, 
{\it decreases very quickly} 
for fluxes away from 
$\Phi_{1}$. 
Thus, the desired peak in 
$J_{c}(\Phi)$ 
is 
{\it too narrow} 
and not very robust against changes in 
$\Phi$. 
It would be greatly desirable 
to have samples with APS with {\it many} periods. 
This multiple-period APS sample would provide 
either very many peaks or an extremely broad peak in $J_{c}(\Phi)$, 
as opposed to just one (narrow) main peak (and its harmonics). 
We achieve this goal (a very broad $J_{c}(\Phi)$) here by studying 
samples with many built-in periods. 

The development of new fabrication technologies 
for pinning arrays with controllable parameters 
allows to fabricate 
not only periodic (square or triangular)
but also 
more complicated 
{\it quasiperiodic } (QP) 
arrays of pinning sites, 
including 
{\it Penrose lattices}
\cite{penrose,debruijn,bookqc}. 

The investigation of physical properties of QP systems 
has attracted considerable interest 
including issues such as band structure and localization of electronic states 
in two-dimensional (2D) Penrose lattice 
\cite{zia,kohmoto}, 
electronic and acoustic properties of one-dimensional (1D) QP lattices
\cite{fnphon1d,fnniu86}, 
superconducting-to-normal phase boundaries of 2D QP micronetworks 
\cite{behrooz,fnniu87,fnniu88,niufn89,linfn02}, 
QP semiconductor heterostructures
and optical superlattices
\cite{zhu}, 
soliton pinning by long-range order 
\cite{kivshar}
and pulse propagation 
\cite{torres}
in QP systems. 
Moreover, 
increasing and, more generally, controlling 
the critical current in superconductors 
by its patterning (perforation)
can be of practical importance for applications
in micro- and nanoelectronic devices. 

The original tiling has been studied in Ref.~\cite{penrose}.
The inflation (or production) rules of
``finite Penrose patterns''
generated by repeated application of deflation and rescaling
have been found, 
which show a definite hierarchical structure 
of the Penrose patterns \cite{bookqc}. 

The electronic and acoustic properties of a one-dimensional quasicrystal 
have been studied in 
Refs.~\cite{kohmoto,fnphon1d}. 
It has been shown, 
in particular, 
that there exist two types of the wave functions, 
self-similar (fractal) and non-self-similar (chaotic)
which show ``critical'' or ``exotic'' behavior
\cite{kohmoto}. 
By both numerical (non-perturbative) and analytical (perturbative)
approaches, 
it has been demonstrated 
\cite{fnphon1d,fnniu86} 
that the phonon and electronic spectra of 1D quasicrystals 
exhibit a self-similar hierarchy of gaps and localized states in the gaps. 
The existence of gaps, and gap states, 
in QP GaAs-AlGaAs superlattices has been predicted and found experimentally. 

Along with studying the structural, electronic and acoustic 
properties of QP structures, 
considerable progress has been reached in understanding 
the superconducting properties of 
2D quasicrystalline arrays 
\cite{behrooz,fnniu87,fnniu88,niufn89,linfn02}. 
The effect of frustration, induced by a magnetic field, 
on the superconducting diamagnetic properties 
has been revealed 
and the superconducting-to-normal phase boundaries, 
$T_{c}(H)$, 
have been calculated for several geometries 
with quasicrystalline order, 
in a good agreement with experimentally measured
phase boundaries
\cite{fnniu87,niufn89}. 
A comprehensive analysis 
of superconducting wire networks including 
quasicrystalline geometries 
and Josephson-junction arrays in a magnetic field 
has been presented in 
Refs~\cite{niufn89,linfn02}. 
An analytical approach 
\cite{niufn89,linfn02} 
was introduced to analyze the structures which are present 
in phase diagrams for a number of geometries. 
It has been shown 
that the gross structure is determined by the statistical distributions
of the cell areas, 
and that the fine structures are determined by correlations among 
neighboring cells in the lattices. 
The effect of thermal fluctuations on the structure of the phase diagram 
has been studied 
\cite{niufn89} 
by a cluster mean-field calculation 
and using real-space renormalization group. 

In this paper, 
we study another phenomenon related to superconducting properties of 
quasiperiodic systems, 
namely, 
vortex pinning 
by 1D QP chains 
and 
by 2D arrays of pinning sites 
located at the nodes of QP lattices 
(e.g., a five-fold Penrose lattice). 
It should be noted that in superconducting networks 
the areas of the network plaquettes play a dominant role 
\cite{behrooz,fnniu87,fnniu88,niufn89}. 
However, 
for vortex pinning by QP pinning arrays, 
the specific geometry of the elements which form the QP lattice 
and their arrangement (and not just the areas) are important, 
making the problem complicated. 

In Sec.~II
we introduce the model used for describing vortex dynamics 
in QP pinning arrays and for determining the critical depinning current, 
$J_{c}$, 
which is analyzed for different quasicrystalline geometries. 

The pinning of vortices by a 1D QP chain of pinning sites
(i.e., Fibonacci sequence) is discussed in Sec.~III. 
We consider a continuous transition from a periodic 
to the QP chain of pinning sites and we monitor the 
corresponding changes in the critical current 
$J_{c}$
as a function of the applied magnetic flux 
$\Phi$. 
A remarkable {\it self-similarity} of 
$J_{c}(\Phi)$ 
is demonstrated in both real space and in reciprocal $k$-space. 

Section~IV studies 
the pinning of flux lattices by 2D QP pinning arrays 
including the five-fold Penrose lattice. 
We analyze changes of the function 
$J_{c}(\Phi)$
during a continuous transition from
a periodic triangular lattice of pinning sites to 
the Penrose lattice. 
Based on detailed considerations of the 
structure and specific local rules of construction 
of the Penrose lattice, 
we predict the main features of the function 
$J_{c}(\Phi)$. 
Numerical simulations with different finite-size Penrose lattices 
confirm the predicted main features for large-size lattices, 
that is important for possible experimental observations 
of the revealed quasiperiodic features. 
We also obtain analytical results supporting our conclusions. 
Moreover, 
we also discuss the changes in the critical current 
by adding either a ``quasiperiodic'' modulation or random displacements 
to initially periodic pinning arrays.

\section{Model}

We model a three-dimensional (3D) slab infinitely long in the $z$-direction, 
by a two-dimensional (2D) (in the $xy$-plane) simulation cell with periodic 
boundary conditions, 
assuming the vortex lines are parallel to the cell edges. 
To study the dynamics of moving vortices 
driven by a Lorentz force, 
interacting with each other
and 
with pinning centers, 
we perform simulated annealing simulations by numerically integrating 
the overdamped equations of motion (see, e.g., 
Ref.~\cite{md01,md0157,md02,md03Z}):
\begin{equation}
\eta {\rm \bf v}_{i} \ = \ {\rm \bf f}_{i} \ = \ {\rm \bf f}_{i}^{vv} + 
{\rm \bf f}_{i}^{vp} + {\rm \bf f}_{i}^{T} + {\rm \bf f}_{i}^{d}. 
\label{eqmd} 
\end{equation} 
Here, 
${\rm \bf f}_{i}$ 
is the total force per unit length acting on vortex 
$i$, 
${\rm \bf f}_{i}^{vv}$ 
and 
${\rm \bf f}_{i}^{vp}$ 
are the forces due to vortex-vortex and vortex-pin interactions, respectively, 
${\rm \bf f}_{i}^{T}$ 
is the thermal stochastic force, 
and 
${\rm \bf f}_{i}^{d}$ 
is the driving force acting on the $i$-th vortex; 
$\eta$ is the viscosity, which is set to unity. 
The force due to the interaction of the $i$-th vortex with other vortices is 
\begin{equation}
{\rm \bf f}_{i}^{vv} \ = \ \sum\limits_{j}^{N_{v}} \ f_{0} \ K_{1} \!
\left( \frac{ \mid {\rm \bf r}_{i} - {\rm \bf r}_{j} \mid }{\lambda} \right)
\hat{\rm \bf r}_{ij} \; , 
\label{fvv}
\end{equation}
where 
$N_{v}$ 
is the number of vortices, 
$K_{1}$
is a modified Bessel function, 
$\lambda$ 
is the magnetic field penetration depth, 
$\hat{\rm \bf r}_{ij} = ( {\rm \bf r}_{i} - {\rm \bf r}_{j} )
/ \mid {\rm \bf r}_{i} - {\rm \bf r}_{j} \mid,$
and 
$$ 
f_{0} = \frac{ \Phi_{0}^{2} }{ 8 \pi^{2} \lambda^{3} } \; . 
$$ 
Here $\Phi_{0} = hc/2e$ is the magnetic flux quantum. 
It is convenient, following the notation used in 
Ref.~\cite{md01,md0157,md02,md03Z}, 
to express all the lengths in units of 
$\lambda$ 
and all the fields in units of 
$\Phi_{0}/\lambda^{2}$. 
The Bessel function 
$K_{1}(r)$ 
decays exponentially for 
$r$
greater than
$\lambda$, 
therefore it is safe to cut off the (negligible) force for distances greater than 
$5\lambda$. 
The logarithmic divergence of the vortex-vortex interaction forces for 
$r \to 0$
is eliminated by using a cutoff for distances less than
$0.1\lambda$. 

Vortex pinning is modeled by short-range parabolic potential wells 
located at positions 
${\rm \bf r}_{k}^{(p)}$. 
The pinning force is 
\begin{equation}
{\rm \bf f}_{i}^{vp} = \sum\limits_{k}^{N_{p}} \left( \frac{f_{p}}{r_{p}} \right)
\mid {\rm \bf r}_{i} - {\rm \bf r}_{k}^{(p)} \mid
\Theta \!
\left( 
\frac{r_{p} - \mid {\rm \bf r}_{i} - {\rm \bf r}_{k}^{(p)} \mid}{\lambda} 
\right)
\hat{\rm \bf r}_{ik}^{(p)},
\label{fvp}
\end{equation}
where 
$N_{p}$
is the number of pinning sites,
$f_{p}$
is the maximum pinning force of each potential well, 
$r_{p}$
is the range of the pinning potential, 
$\Theta$ 
is the Heaviside step function, 
and 
$\hat{\rm \bf r}_{ik}^{(p)} = ( {\rm \bf r}_{i} - {\rm \bf r}_{k}^{(p)} )
/ \mid {\rm \bf r}_{i} - {\rm \bf r}_{k}^{(p)} \mid.$

The temperature contribution to Eq.~(\ref{eqmd}) is represented by a stochastic 
term obeying the following conditions: 
\begin{equation}
\langle f_{i}^{T}(t) \rangle = 0
\end{equation}
and
\begin{equation}
\langle f_{i}^{T}(t)f_{j}^{T}(t^{\prime}) \rangle = 2 \,  \eta \,  k_{B} \,  T \,  \delta_{ij} \,  \delta(t-t^{\prime}). 
\end{equation}

The ground state of a system of moving vortices is obtained as follows. 
First, we set a high value for the temperature, to let vortices move randomly. 
Then, the temperature is gradually decreased down to $T = 0$. 
When cooling down, vortices interacting with each other and 
with the pinning sites adjust themselves to minimize the energy, 
simulating the field-cooled experiments \cite{tonomura-vvm,togawa}. 

In order to find the critical depinning current, 
$J_{c}$,
we apply 
an external driving force gradually increasing from 
$f_{d} = 0$
up to a certain value 
$f_{d} = f_{d}^{c}$,
at which all the vortices become depinned 
and start to freely move. 
For values of the driving force just above 
$f_{d}^{c}$, 
the total current of moving vortices 
$J \sim \langle v \rangle$ becomes nonzero. 
Here, 
$\langle v \rangle$ 
is the normalized (per vortex) 
average velocity of all the vortices 
moving in the direction of the applied driving force. 
In numerical simulations, this means that, in practice, one should define 
some threshold value 
$J_{\rm min}$
larger than the noise level. 
Values larger than 
$J_{\rm min}$
are then considered as nonzero currents. 
However, instead 
using this criterion-sensitive scheme, 
we can use an alternative approach based on potential energy considerations. 
In case of deep short-range 
potential wells, the energy required to depin vortices trapped by 
pinning sites is proportional to the number of pinned vortices, 
$N_{v}^{(p)}$. 
Therefore, 
in this approximation 
we can define the ``critical current'' as follows: 
\begin{equation}
J_{c}(\Phi) = J_{0} \frac{N_{v}^{(p)}(\Phi)}{N_{v}(\Phi)},
\label{jc}
\end{equation}
where 
$J_{0}$
is a constant, 
and study the dimensionless value
$J_{c}^{\prime} = J_{c}/J_{0}$
(further on, the primes will be omitted). 
Throughout this work, we use narrow potential wells as pinning sites, 
characterized by 
$r_{p} = 0.04\lambda$ to $0.1\lambda$. 
Our calculations show that, for the parameters used, the dependence 
of the critical current 
$J_{c}(\Phi)$
defined according to Eq.~(\ref{jc}), 
is in good agreement to that based on the above general 
definition of 
$J_{c}$ 
(which involves an adjustable parameter $J_{\rm min}$). 
The advantages of using 
$J_{c}$ 
defined by Eq.~(\ref{jc}) are the following: 
it 
(i) does not involve any arbitrary threshold 
$J_{\rm min}$ 
and 
(ii) is less CPU-time consuming, allowing the study of very large-size lattices. 
Moreover, the goal of this study is to reveal specific matching effects between 
a (periodic) vortex lattice and arrays of QP pinning sites, 
and to study how the quasiperiodicity manifests itself in experimentally measurable quantities 
($J_{c}$, $N_{v}^{(p)}/N_{v}$) 
related to the vortex pinning by QP (e.g., the Penrose lattice) pinning arrays.

\section{Pinning of vortices by a 1D quasiperiodic chain of pinning sites}

In this section we study the pinning of vortices by one dimensional (1D) QP 
chains of pinning sites.

\subsection{1D quasiperiodic chain}

As an example of a 1D QP chain, or 1D 
{\it quasicrystal}, 
a Fibonacci sequence is considered, which can be constructed following a 
simple procedure: 
let us consider two line segments, long and short, denoted, respectively, by 
$L$ and $S$. 
If we place them one by one, we obtain an infinite periodic sequence: 
\begin{equation}
LSLSLSLSLSLSLSLS \, ... 
\label{ls}
\end{equation}
A unit cell of this sequence consists of two elements, 
$L$ and $S$. 
In order to obtain a QP sequence, these elements are transformed
according to 
Fibonacci rules
as follows: 
$L$ is replaced by $LS$, 
$S$ is replaced by $L$:
\begin{equation}
L \rightarrow LS, \ \ \ S \rightarrow L.
\label{rule}
\end{equation}
As a result, we obtain a new sequence: 
\begin{equation}
LSLLSLLSLLSLLSL \, ... 
\label{lsl}
\end{equation}

Iteratively applying the rule (\ref{rule}) to the sequence (\ref{lsl}), 
we obtain, 
in the next iteration, a sequence with a five-element unit cell 
$(LSLLS)$, 
then with an eight-element unit cell 
$(LSLLSLSL)$, 
and so on, to infinity. 
For the sequence with an $n$-element cell, where $n \to \infty$, 
the ratio of numbers of long to short elements is the golden mean value, 
\begin{equation}
\tau = (1 + \sqrt{5})/2.
\label{golden}
\end{equation}
The position of the $n$th point where a new element, 
either $L$ or $S$, begins 
is determined by 
\cite{bookqc}: 
\begin{equation} 
x_{n} = n + \gamma \left[ \gamma n \right], 
\label{qpcoor} 
\end{equation} 
where 
$\left[ x \right]$
denotes the maximum integer less or equal to 
$x.$
Equation (\ref{qpcoor}) corresponds to the case when 
the Fibonacci sequence has a ratio 
$\gamma = a_{S}/a_{L}$ 
of the length of the short segment, 
$a_{S}$,
to the length of the long segment, 
$a_{L}$. 
Ratios 
$\gamma$
other than 
$1 / \tau$
correspond to other chains which are all QP. 
Along with the golden mean value of 
$\gamma = 1 / \tau$, 
we use in our simulations 
$\gamma$'s
varying in the interval between 0 and 1:
$0 < \gamma < 1$,
when analyzing a continuous transition from a periodic
to a QP (Fibonacci sequence) pinning array. 

To study the critical depinning current $J_{c}$ in 1D QP 
pinning chains, we place pinning sites on the points where 
$L$ or $S$
elements of the QP sequence link to each other.
Therefore, the coordinates of the centers of the pinning sites 
are defined by Eq.~(\ref{qpcoor}) with 
$\gamma = a_{S}/a_{L}$. 

\begin{figure}[btp]
\begin{center}
\vspace*{-2.5cm} 
\hspace{-1.0cm}
\includegraphics*[width=9.5cm]{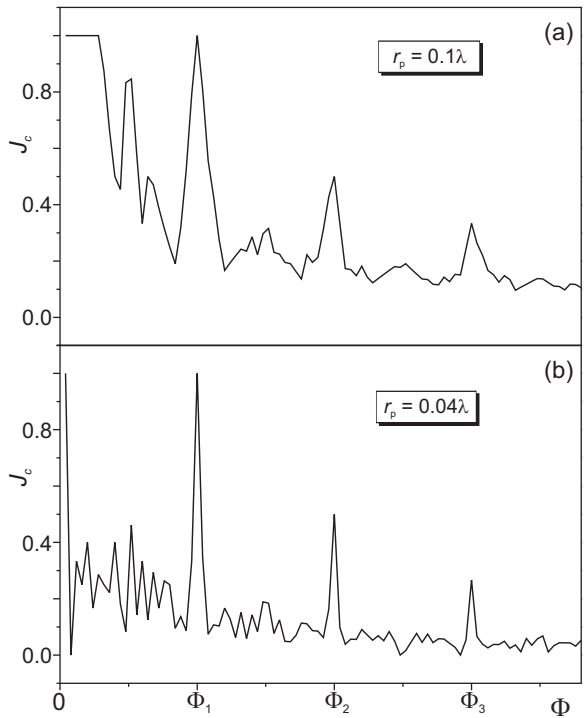}
\end{center}
\vspace{-1.0cm} 
\caption{ 
(a, b)
Dimensionless critical depinning current 
$J_{c}$, 
as a function of the applied 
magnetic flux 
$\Phi$, 
in a 1D periodic chain of pinning sites, for a
cell containing 25 pinning sites, 
$N_{p} = 25$ 
and 
$f_{p}/f_{0} = 2.0$. 
The indicated fluxes 
$\Phi_{1}$, 
$\Phi_{2}$, and 
$\Phi_{3}$ 
correspond 
to the first, second, and third matching fields.
The function $J_{c}(\Phi)$ 
is shown 
for two different
values of the pinning site radius, 
$r_{p} = 0.1\lambda$ (a), 
and
$r_{p} = 0.04\lambda$ (b). 
When 
$r_{p}$, 
decreases, 
the 
main commensurability peaks become sharper, 
as shown in (a) and (b). 
}
\end{figure}

\subsection{Pinning of vortices by a 1D periodic chain of pinning sites}

We start with a periodic chain of pinning sites, 
which can be considered, 
in the framework of the above scheme, 
as a limiting case of a ``QP'' chain 
with 
$\gamma = 1$,
i.e.
$a_{S} = a_{L} = 1$.
In Fig.~1a, the critical current 
$J_{c} \sim N_{v}^{(p)}/N_{v}$ 
is shown as a function of the applied magnetic flux 
$\Phi$ 
for 
$f_{p}/f_{0} = 2$
and for a pinning site radius 
$r_{p} = 0.1 \lambda$.
Sharp peaks of the function 
$J_{c}(\Phi)$ 
correspond to matching fields. 
Since
the dimensionless critical current we plot 
is that per vortex and is proportional 
to the number of pinned vortices and inversely proportional 
to the total number of vortices (Eq~(\ref{jc})),
therefore, the magnitude of the peaks versus 
$\Phi \sim N_{v}$
decreases as 
$1/\Phi$. 
Then 
the maximum heights of the peaks are:
$J_{c}(\Phi_{1}) = 1$, 
$J_{c}(2\Phi_{1}) = 0.5$, 
and
$J_{c}(3\Phi_{1}) = 0.33$.
%
%
Note that these values are obtained provided 
each pinning site can trap only one vortex, 
which is justified 
for the chosen radius of the pinning site and for the vortex densities considered. 
In addition, 
there are weak wide maxima corresponding to 
$\Phi_{1}/2$ 
and other ``subharmonics'', 
i.e. 
$\Phi_{i} + \Phi_{1}/2$ (Fig.~1a),
where 
$i$ 
is an integer. 
For very small values of 
$\Phi_{1}$,
the 
vortex density is very low, 
and vortices almost do not interact with each other. 
As a result, in the ground state they all become trapped by pinning sites, 
and 
$J_{c}$
is maximal for small
$\Phi_{1}$. 

For smaller radii of the pinning sites, 
$r_{p} = 0.04 \lambda$ (Fig.~1b), 
the 
$J_{c}(\Phi)$ 
peaks corresponding to matching fields become sharper
because for smaller values of 
$r_{p}$,
it is more difficult to fulfill the commensurability conditions. 
Any features 
around the main peaks, 
including subharmonics, 
are suppressed. 
Note also a ``parity effect'' takes place in this case: 
since the number of pinning sites per cell is odd 
($N_{p} = 25$ in Fig.~1),
therefore 
$J_{c}(\Phi)$ 
peaks 
are suppressed
for 
$\Phi_{1}/2$
and for 
{\it even} 
values of 
$i$ in the sequence 
$\Phi_{i} + \Phi_{1}/2$.

\begin{figure}[btp]
\begin{center}
\vspace*{-1.0cm} 
\hspace{-2.0cm}
\includegraphics*[width=18.0cm]{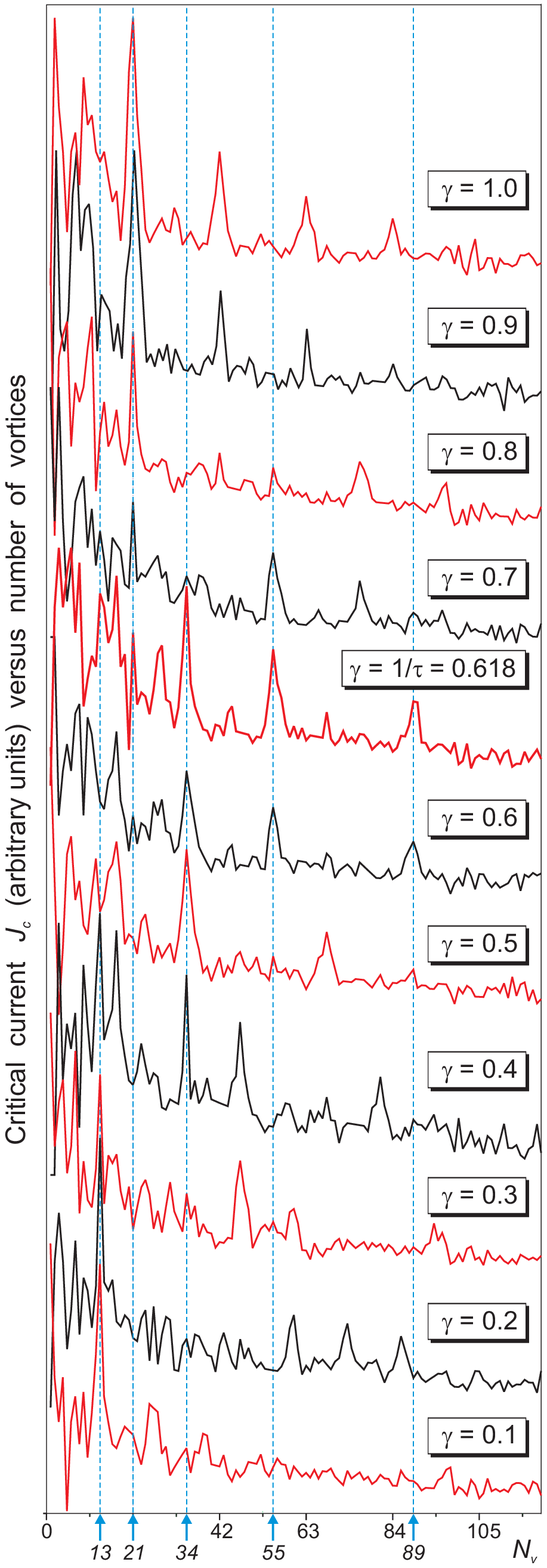}
\end{center}
\vspace{-1.0cm} 
\nonumber
\end{figure}

\begin{figure}[btp]
\caption{ 
(Color)
Critical depinning current 
$J_{c}$, 
as a function of the 
applied magnetic flux 
(for convenience, shown as a function of the number
of vortices, 
$N_{v} \sim \Phi$), 
in a 1D QP chain of pinning sites
with
$f_{p}/f_{0} = 1$,
$r_{p} = 0.1\lambda$, 
for chains characterized by different ratios 
of the lengths of short 
$a_{S}$ 
to long 
$a_{L}$ 
spacings, 
$\gamma = a_{S}/a_{L}$. 
For small deviations of the chain from a periodic chain 
($\gamma = 0.9$ to 1), 
commensurability peaks are similar to those shown in Fig.~1. 
For intermediate values of 
$\gamma$
($\gamma = 0.2$ to 0.8), 
peaks are determined by a sequence of harmonics of numbers of long and short 
periods of the chain, 
which includes the sequence of successive Fibonacci numbers, most pronounced for 
$\gamma = 1 / \tau \approx 0.618$, 
where 
$\tau = (1 + \sqrt{5})/2 \approx 1.618$ is the golden mean. 
For very small 
$\gamma$ 
(e.g., $\gamma = 0.1$), 
the QP chain effectively becomes periodic
but with the number of pinning sites equal to the number
of long periods in the chain. 
}
\end{figure}

\section{Pinning of vortices by a 1D quasiperiodic chain of pinning sites:
Gradual evolution from a periodic to a quasiperiodic chain}


Let us consider a QP chain of pinning sites
with 
spacings between pinning sites given by 
$a_{L}$ (long)
and 
$a_{S}$ (short).
The long and short segments alternate according to the Fibonacci rules
forming a Fibonacci sequence. 
The number of pinning sites per cell 
coincides with 
the number of elements of this sequence per unit cell.
It is natural 
to take 
a chain with a number of elements equal to 
one of the successive Fibonacci numbers 
as a 1D cell, 
although in principle it could be of any length. 
We impose periodic boundary conditions at the ends of the cell.

The larger cell we take, 
the closer we are to describing a truly QP structure. 
However, it turns out, 
that even a 
finite part 
of a QP system (1D chain or 2D QP lattice) 
provides us with reliable information
concerning properties of the 
whole 
system. 
This is based on the 
structural 
{\it self-similarity }
of QP systems. 
These properties are studied here for the critical depinning current
$J_{c}$ 
and have also been demonstrated for other physical phenomena 
\cite{fnphon1d,fnniu86,fnniu87,fnniu88,niufn89,linfn02,kolarfn90}.

Figure~2 shows 
the evolution of 
$J_{c}$ 
as a function of 
the number of vortices, 
$N_{v} \sim \Phi$, 
for various values of the parameter
$\gamma$.
The top curve represents the limiting case of a periodic chain
($\gamma = 1.0$)
with typical peak structure discussed above (Fig.~1).
The chain contains 
$N_{p} = 21$
pinning sites. 
As a result, commensurability peaks appear at 
$N_{v} = 21,$ 42, 63, etc., i.e. multiples of 21. 
A small QP 
distortion 
of the chain
does not appreciably affect the peak structure of the
function 
$J_{c}(N_{v})$ 
($\gamma = 0.9$). 
When the deviation of the factor 
$\gamma$ 
from unity becomes larger
($\gamma = 0.8$, 0.7), 
commensurability peaks for 
$N_{v} = 42,$ 63, and other multiples of 21
decrease in magnitude. 
At the same time, 
new peaks appear at
$N_{v} = 55,$ 76 ($= 55 + 21$), 97 ($= 76 + 21$)
for 
$\gamma = 0.8$. 
Then, 
with further decrease of 
$\gamma$
($\gamma = 0.7$), 
these peaks remain 
($N_{v} = 55$ 
even grows in magnitude), 
and a new peak at 
$N_{v} = 34$ 
arises. 

For the golden mean-related value of 
$\gamma = 1/\tau \approx 0.618$, 
we obtain a set of peaks, which are ``harmonics'' of 
the numbers of long and short periods of the chain
(or reciprocal lengths $a_{L}$ and $a_{S}$), 
i.e. 
\begin{equation}
x_{{\rm peaks},\ i}^{QP} \ = \ A_{i}N_{L} + B_{i}N_{S} \ = \ \frac{A_{i}^{\prime}}{a_{S}} + \frac{B_{i}^{\prime}}{a_{L}}, 
\label{xpeaks}
\end{equation}
where 
$N_{L}$
and 
$N_{S}$
are the numbers of long and short elements, respectively;
$A_{i}$ $(A_{i}^{\prime})$ 
and 
$B_{i}$ $(B_{i}^{\prime})$ 
are generally (positive or negative) multiples or divisors of 
$N_{L}$
and 
$N_{S}$;
the upper index ``QP'' denotes ``quasiperiodic''.
It is easy to see that  
this set includes as a subset the 
sequence of successive Fibonacci numbers. 
In particular, the following 
well-resolved 
peaks of the function 
$J_{c}(N_{v})$ 
appear for 
$\gamma = 1/\tau \approx 0.618$:
$N_{v} = 13$ ($= N_{L}= 13$, 13 is a Fibonacci number (FN));
$N_{v} = 17$ ($= (2N_{L} + N_{S})/2$, where $(2N_{L} + N_{S}) = 34$ is a FN);
$N_{v} = 21$ ($= N_{p} = N_{L} + N{S}$, 21 is a FN);
$N_{v} \approx 27$-28 ($= (3N_{L} + 2N_{S})/2$, where $(3N_{L} + 2N_{S}) = 55$ is a FN); 
$N_{v} = 34$ $(2N_{L} + N_{S})$; 
$N_{v} \approx 44$-45 ($= (5N_{L} + 3N_{S})/2$, where $(5N_{L} + 3N_{S}) = 89$ is a FN); 
$N_{v} = 55$ ($= (3N_{L} + 2N_{S})$, 55 is a FN); 
$N_{v} = 68$ ($= (4N_{L} + 2N_{S})$);
$N_{v} = 89$ ($= (5N_{L} + 3N_{S})$, 89 is a FN). 
In summary, 
the most pronounced peaks are at: 
$N_{v} = 13,$ 17, 21, 34, 55, 89, 
which (except the point $N_{v} = 17 = 34/2$) form a sequence of successive FNs.

\begin{figure}[btp]
\caption{ 
(Color)
(a) The critical depinning current 
$J_{c}$, 
as a function of 
the number of vortices, 
$N_{v} \sim \Phi$, 
for different 1D QP chains, 
$N_{p} = 21$ (red bottom line),
$N_{p} = 34$ (blue line), 
$N_{p} = 55$ (green line),
and
$N_{p} = 89$ (dark blue top line), 
and the same
$\gamma = a_{S}/a_{L} = 1 / \tau$. 
The parameters used here are: 
$f_{p}/f_{0} = 1.0$ 
and 
$r_{p} = 0.1\lambda$. 
Independently of the length of the chain, 
the peaks for all of the curves include as a subset 
the sequence of successive Fibonacci numbers 
(indicated by blue arrows in the horizontal axis) 
and their sub-harmonics. 
(b) $J_{c}$($N_{v}$) 
for a long chain 
$N_{p} = 144$ 
and the same
$\gamma = 1 / \tau$. 
Notice that now $N_{v}$ is much larger than in (a). 
(c) The function 
$J_{c}(\Phi/\Phi_{1})$
for the same set of 1D chains (using the same colors), 
normalized by the number of pinning sites for each chain. 
The curves for chains with different $N_{p}$'s display the same sets of peaks, 
namely, 
at 
$\Phi/\Phi_{1}=1$ (first matching field)
and 
$\Phi/\Phi_{1}=0.5$, 
as well as
at the golden-mean-related values: 
$\Phi/\Phi_{1}=\tau/2$,
$\Phi/\Phi_{1}=(\tau+1)/2=\tau^{2}/2$,
$\Phi/\Phi_{1}=\tau$,
$\Phi/\Phi_{1}=(\tau+\tau^{2})/2=\tau^{3}/2$,
$\Phi/\Phi_{1}=\tau^{2}=\tau+1$,
$\Phi/\Phi_{1}=\tau^{2}+1$. 
Therefore, similar sets of peaks are obtained for both cases:
for the curves plotted in the same scale (a), (b);
for the curves plotted in individual scales, 
i.e., 
normalized on the number of pinning sites in the chain (c). 
This behavior demonstrates the self-similarity of $J_{c}(\Phi)$. 
}
\end{figure}

\begin{figure}[btp]
\begin{center}
\vspace*{-0.5cm} 
\hspace*{-3.0cm}
\includegraphics*[width=14.5cm]{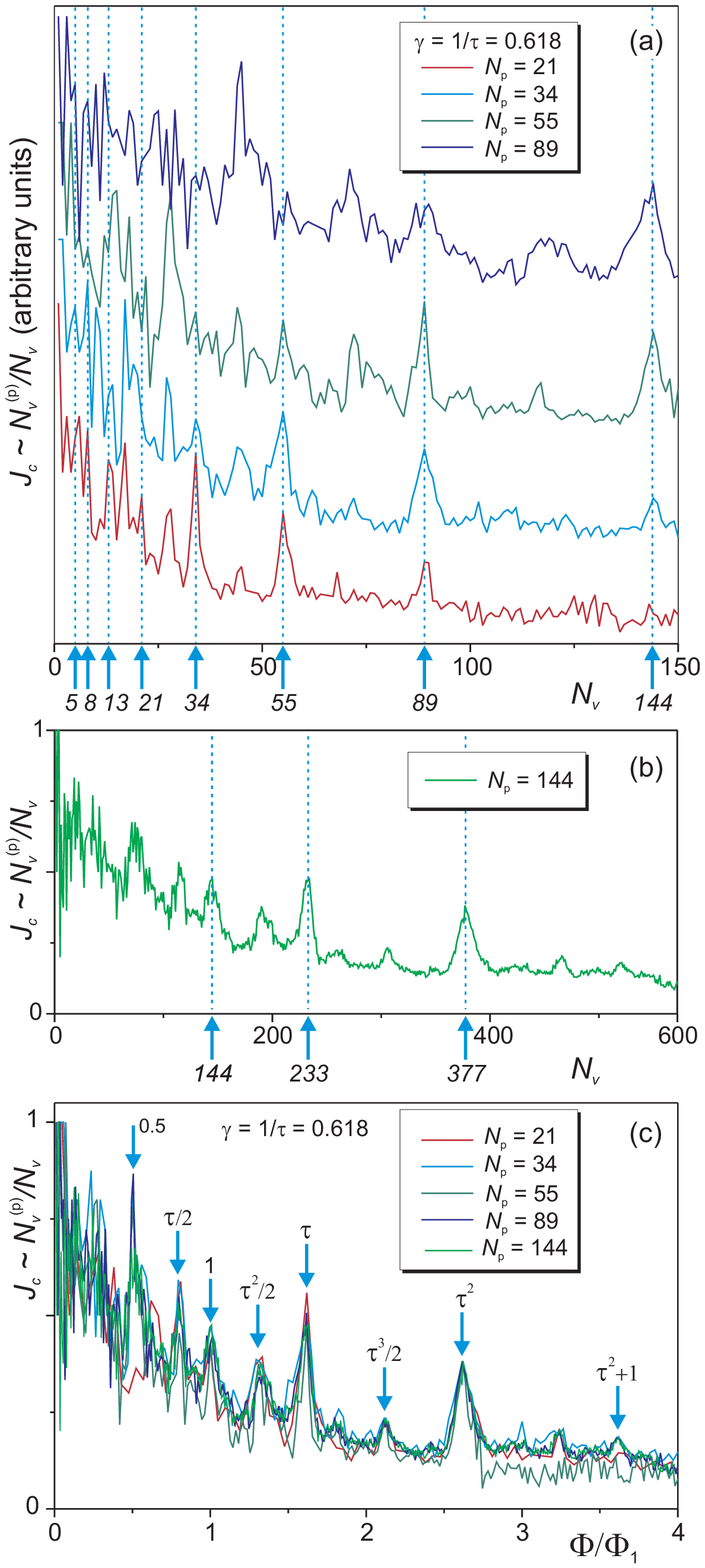}
\end{center}
\vspace{-2.0cm} 
%
\nonumber
\end{figure}

The above QP peaks only slightly degrade at 
$\gamma = 0.6$. 
However, when the length of the long segment 
$a_{L}$
becomes twice the length of the short one 
$a_{S}$, 
i.e. 
$\gamma = 0.5$, 
sharp commensurability peaks appear which are related to the small segment of the chain 
with length 
$a_{S} = a_{L}/2$. 
Namely, 
we obtain peaks at 
$N_{v} = 34 \sim (2N_{L} + N_{S})$
and at other values of 
$N_{v}$
which are 
sub-harmonics 
of $2N_{L} + N_{S}$:
$N_{v} = 17,$ 34, 68. 
Other peaks, in particular, those related to the Fibonacci sequence, 
are much less pronounced
for $\gamma = 0.5$. 
For 
$\gamma = 0.4$, 
the peaks are at: 
$N_{v} = 13,$ 17, 34, 47 ($= 3N_{L} + N_{S}$), 68, and 81. 
A very strong (recall that the maximum amplitude of the peak is $\sim 1/N_{v}$) peak at 
$N_{v} = 81$
is a ``resonance'' peak corresponding to the ratio of
$B_{i}/A_{i} = 2/5 = \gamma$, 
i.e.
$5N_{L} + 2N_{S} = 81$. 
When 
$\gamma = 0.3$, 
the ``resonance'' ratio
$B_{i}/A_{i} = 0.3$, 
therefore, 
a strong peak appears for the ``nearest to 0.3'' value
$B_{i}/A_{i} = 0.33:$
$3N_{L} + N_{S} = 47$, 
and also for 
$6N_{L} + 2N_{S} = 94$. 
Another ``close to 0.3'' value is
$B_{i}/A_{i} = 0.25,$
which is responsible for the peaks at 
$N_{v} = 30$ ($2N_{L} + N_{S}/2$)
and 
$N_{v} = 60$ ($4N_{L} + N_{S}$). 
Also, peaks at
$N_{v} = 13,$ 17 and 55
are present. 
The ``resonant'' peak for 
$\gamma = 0.2$
(i.e., for the ratio 
$\gamma = B_{i}/A_{i} = 0.2$),
appears at 
$N_{v} = 73$ ($= 5N_{L} + N_{S}$).
The closest neighboring peaks are at
$N_{v} = 60$ ($= 4N_{L} + N_{S}$)
and 
$N_{v} = 86$ ($= 6N_{L} + N_{S}$), 
[also: $N_{v} = 43$ ($= 3N_{L} + N_{S}/2$)], 
characterized by ratios
$B_{i}/A_{i} = 0.25$
and
$B_{i}/A_{i} = 0.17,$
correspondingly. 
Finally, for 
$\gamma = 0.1$
we arrive at the situation when we have an almost periodic chain
but with a period 
$a$
different from that for 
$\gamma = 1.0$:
$a^{\prime} \approx a_{L}$ since 
$a_{S} \ll a_{L}.$
The number of pinning sites becomes
$N_{p} = N_{L} = 13$, 
and we obtain commensurability peaks at
the positions which are 
multiples of 
$N_{p} = 13$, 
i.e., at 
$N_{v} = 13,$ 26, 39, 52, 
which is typical for periodic chains of pinning sites. 

In order to demonstrate that the above analysis is general 
and reveals the QP features 
independently of the length of a specific chain of pinning sites, 
let us compare results 
from the calculation of 
$J_{c}(N_{v})$
for different chains. 
In Fig.~3a, 
the function 
$J_{c}(N_{v})$
is shown for 
four different 1D QP chains, 
$N_{p} = 21$,
$N_{p} = 34$, 
$N_{p} = 55$, 
and 
$N_{p} = 89$, 
and the same
$\gamma = 1/\tau$. 
Figure~3a clearly shows that the 
{\it positions} of the {\it main peaks in $J_{c}$}, 
i.e. those corresponding to a Fibonacci sequence, 
and other peaks 
whose positions are described by Eq.~(\ref{xpeaks}), 
to a significant extent, 
do {\it not} depend on the length of the chain.
The peaks 
shown in Fig.~3
form a Fibonacci sequence:
$N_{v} = 13,$ 21, 34, 55, 89, 144, 
and other ``harmonics'':
$N_{v} = 17,$ 27-28 ($= 55/2$), 44-45 ($= 89/2$), etc. 
At the same time, 
longer chains allow to better reveal peaks for larger Fibonacci numbers. 
Thus, for chains with $N_{p} = 144$ (Fig.~3b) peaks at the next Fibonacci numbers 
are pronounced: $N_{v} = 144,$ 233, 377. 

While the curves for different chains are plotted in Fig.~3a in 
the same 
scale, 
Fig.~3c shows these curves in 
individual 
scales. 
Namely, 
we rescale each $J_{c}$ by normalizing each $J_{c}$ by the number of pinning sites 
in each curve. 
Thus, 
$\Phi_{1}$ corresponds to 
$N_{v} = 21$
for 
the chain with
$N_{p} = 21$, 
to
$N_{v} = 34$
for 
the chain with
$N_{p} = 34$, etc. 
After this rescaling, the $J_{c}$ curves approximately follow each other 
and have pronounced peaks for the golden-mean-related values 
of 
$\Phi/\Phi_{1}$, 
as shown in Fig.~3c. 
For example, 
$\Phi/\Phi_{1} = \tau$ 
corresponds 
to 
$N_{v} = 34$
for 
the chain with
$N_{p} = 21$, 
to 
$N_{v} = 55$
for 
the chain with
$N_{p} = 34$, 
to 
$N_{v} = 89$
for 
$N_{p} = 55$, 
to 
$N_{v} = 144$
for 
$N_{p} = 89$, 
and 
to 
$N_{v} = 233$
for 
the chain with
$N_{p} = 144$. 
Note that these peaks 
(i.e., corresponding to the golden mean) 
are most pronounced for each chain in Fig.~3a,b. 

Therefore, 
the same peaks of the function 
$J_{c}(\Phi)$
for different chains 
are revealed before and after rescaling. 
This means that 
the function 
$J_{c}(\Phi)$
for the 1D QP chain 
is 
{\it self-similar}. 
Below, we demonstrate 
the revealed self-similarity effect in 
a reciprocal 
$k$-space.

\subsection{Fourier-transform of the vortex distribution function 
on a 1D periodic chain of pinning sites: Self-similarity effect}

As we established above, 
the revealed QP features 
(e.g., peaks of the function 
$J_{c}(\Phi \sim N_{v})$
) 
(i) are independent of the length of the chain
and 
(ii) the longer chain we take, the more details (e.g., ``subharmonics'') 
of QP features can be observed. 

This result is related to an important property of QP systems, 
self-similarity, 
which could be better understood 
by 
analyzing the Fourier-transform
of the distribution function of the system of vortices pinned on 
a QP array. 

In $k$-space, 
the distribution function 
$F_{v}^{(p)}(q)$
of a system of 
$N_{v}^{(p)}$
pinned vortices can be represented as the inverse Fourier-transform
of the 1D distribution function 
$F_{v}^{(p)}(n)$
of the vortices in real space: 
\begin{equation}
F_{v}^{(p)}(q) = \frac{1}{N_{v}^{(p)}} \sum \limits_{n = 1}^{ N_{v}^{(p)} } 
F_{v}^{(p)}(n) 
\, \exp{ \{ -2 \pi i q n / N_{v}^{(p)} \} }. 
\label{fourier}
\end{equation}

\begin{figure}[btp]
\begin{center}
\vspace*{-0.5cm} 
\hspace*{-2.5cm}
\includegraphics*[width=13.5cm]{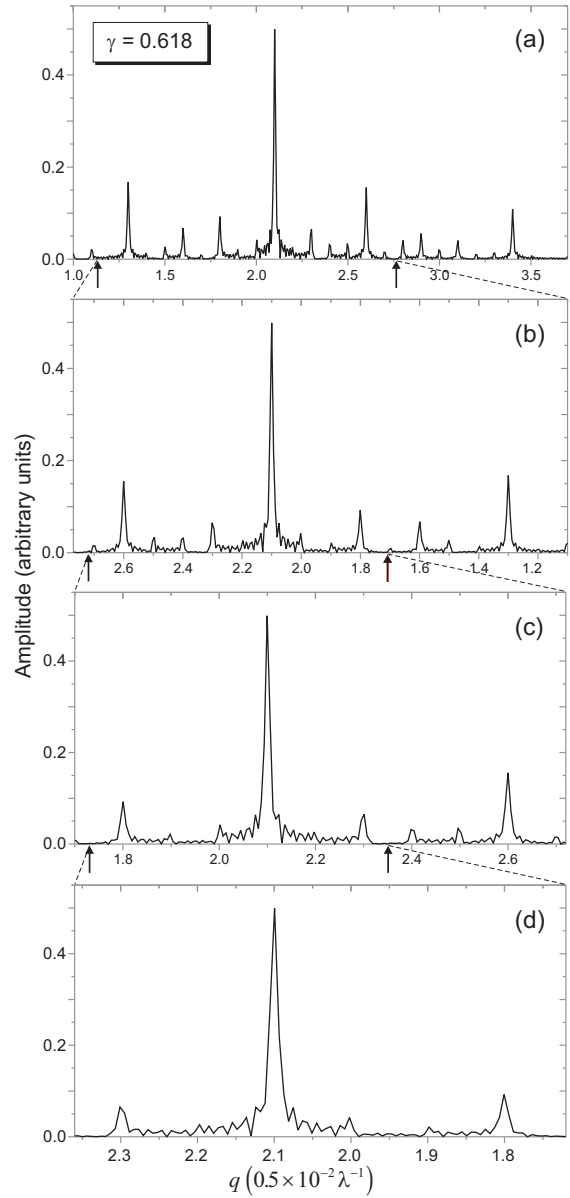}
\end{center}
\vspace{-1.0cm} 
\caption{ 
The self-similar Fourier-transform of the distribution function 
(vortex density) 
of the system of $N_{v} = 144$ vortices pinned on a QP array, 
for 
$\gamma = 1 / \tau$. 
The portion limited by the two arrows in (a) is successively 
magnified 
several times 
and 
the corresponding results 
shown in (b), 
(c) and (d). 
}
\end{figure}

Figure~4 shows 
the Fourier-transform of a system of pinned vortices for a
$\gamma$
equal to the inverse golden mean: 
$1 / \gamma = a_{L}/a_{S} = \tau = (1 + \sqrt{5})/2$. 
The plots shown in Figs.~4a to 4d are obtained according to the 
following rule. 
The portion 
of the horizontal axis which is 
limited by the two arrows in Fig.~4a is rescaled and shown in Fig.~4b.
In the same way, 
the portion limited by the two arrows in Fig.~4b is rescaled and shown in Fig.~4c.
Fig.~4d is obtained following the same procedure. 
Note that each subsequent scaling is accompanied by flipping the 
direction of the $q$-axis to the opposite direction. 
A similar property is clear from the experimental diffraction patterns 
of quasicrystals \cite{bookqc}. 
The pentagons of Bragg

\begin{figure}[btp]
\begin{center}
\vspace*{-0.5cm} 
\hspace*{-1.5cm}
\includegraphics*[width=12.0cm]{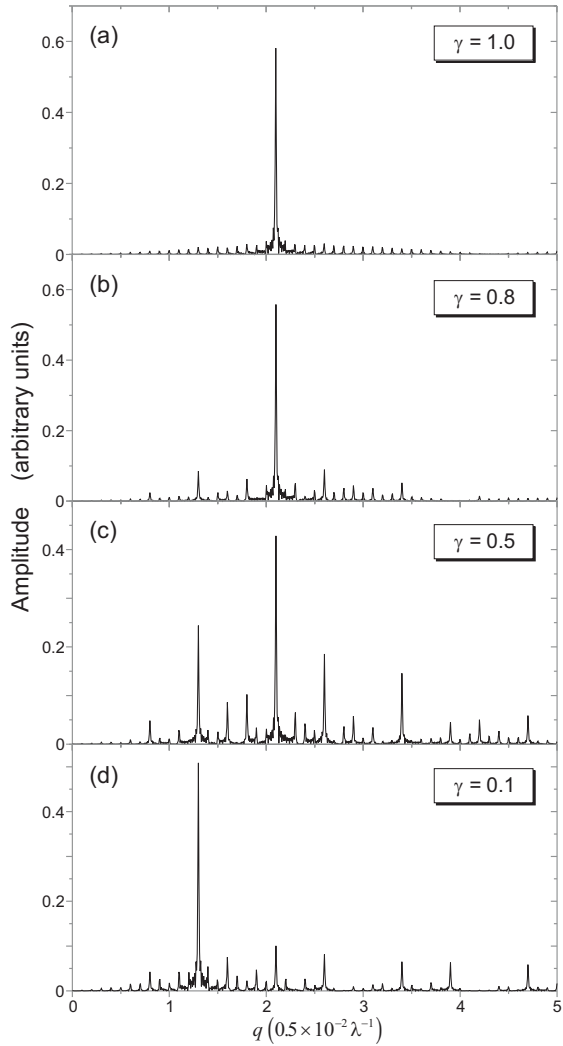}
\end{center}
\vspace{-1.0cm} 
\caption{ 
Fourier-transform of the distribution function 
of the vortices ($N_{v} = 144$) 
interacting with 
a QP array 
of pinning sites, 
for different values of 
$\gamma = 1.0$ (a), 
0.8 (b), 0.5 (c), 0.1 (d). 
The value $\gamma = 1$ corresponds to the limit of a periodic chain. 
Varying $\gamma$ from one, we introduce quasiperiodicity (the most pronounced for $\gamma = 1/\tau \approx 0.618$) 
in the chain. 
The $\gamma = 1$ case recovers the periodic limit (with another period). 
The main central sharp peak, corresponding to the periodic chain used for (a), 
continuously transforms --- through the set of self-similar 
patterns ((b) and (c)) corresponding to QP chains (see Fig.~4) --- 
to another peak (d) produced by a periodic chain with the number
of sites equal to the number of long segments of the initial 
periodic chain with 
$\gamma = 1.0$ 
shown in (a). 
}
\end{figure}

\noindent
peaks have smaller pentagons inside them, 
which are inverted. 
As seen in Fig.~4, 
each subsequent subdivision leads to a subset of peaks 
similar to the entire set of peaks. 

This analysis clearly demonstrates the self-similarity of 
the distribution function of the vortices pinned on a 
QP 1D array of pinning sites. 
Similarly to the observed behavior of the function 
$J_{c}(\Phi)$
when increasing the length of the chain of pinning sites, 
the Fourier-transform of the
distribution function of the vortices 
pinned on a QP array
reproduce 
its main features (peaks)
in a self-similar way,
when increasing the range in $k$-space, 
and simultaneously acquires a more elaborate structure 
with smaller self-similar peaks.

As we discussed above, 
the main commensurability peaks evolve from a perfectly periodic
set of the type
$\Phi_{i} = m \Phi_{1}$, 
where 
$m > 1$
is a positive integer, 
(through the set of QP peaks defined by Eq.~(\ref{xpeaks})), 
to another set of periodic peaks 
$\Phi_{i}^{\prime} = m \Phi_{1}^{\prime}$, 
when the ``quasiperiodicity parameter'' 
$\gamma \equiv a_{S}/a_{L}$ 
is gradually tuned between the values 
$\gamma = 1.0$
and
$\gamma = 0$
(see Fig.~2). 
These limits ($\gamma = 1.0$ and $\gamma = 0$) correspond to 
periodic chains. 

Fig.~5 illustrates the corresponding evolution 
of the Fourier-transform 
of the distribution function of vortices pinned on 
a 1D QP array of pinning sites. 

For 
$\gamma = 1$ (Fig.~5a), 
there is a single sharp peak 
(accompanied by negligibly small satellites)
corresponding to 
a periodic chain. 
For smaller 
$\gamma$'s 
(e.g., $\gamma = 0.8$ (Fig.~5b), 
$\gamma = 0.8$ (Fig.~5c), 
or 
$\gamma = 0.618$ (Fig.~4a)), 
a set of satellite peaks appears around the main peak. 
Simultaneously, 
the intensity of this peak decreases
giving rise to 
another main peak for smaller value of $q$ (Fig.~5d), 
which corresponds to a periodic chain with a larger period.

\section{Pinning of vortices by 2D quasiperiodic pinning arrays}

In the previous section we studied the pinning of vortices by 1D QP 
chains of pinning sites. 
In particular, 
we showed how the quasiperiodicity manifests itself in 
the critical depinning current, 
$J_{c} \sim N_{v}^{(p)}/N_{v}$,
when increasing the applied magnetic flux, 
$\Phi \sim N_{v}$. 
We found that 
the positions of the peaks of the function 
$J_{c}(\Phi)$
are governed by ``harmonics'' of long and short periods 
of the QP chain of pinning sites. 
Independently of the length of the chain
(for $N_{p} \geq 21$), 
the peaks form a QP sequence
including the Fibonacci sequence as a fundamental subset. 
This self-similarity effect is clearly displayed 
in the Fourier-transform of the distribution function
of the vortices on a 1D QP 
array of pinning centers. 
The evolution of QP peaks,
when gradually changing the ``quasiperiodicity'' parameter 
$\gamma$,
has revealed a continuous transition from a QP chain ---
through the set of QP states 
(most pronounced for 
$\gamma = 1/\tau \approx 0.618$) --- 
to another periodic chain, $\gamma = 0$, with a longer period. 
This phenomenon has been studied both in real space 
and in reciprocal $k$-space. 

In the present section
and in the next sections, 
we analyze vortex pinning by 2D QP
arrays; 
in particular, 
by an array of pinning sites placed in the nodes 
of a {\it five-fold Penrose lattice.}
Before tackling the Penrose-lattice pinning array, 
let us start with 
a simplified system
which one can call
``2D-quasiperiodic'' (2DQP)
since it is a 2D system
periodic in one direction
($x$-direction)
and 
QP

\begin{widetext}

\begin{figure}[btp]
\begin{center}
\vspace*{-1.5cm} 
\hspace*{-1.0cm}
\includegraphics*[width=19.5cm]{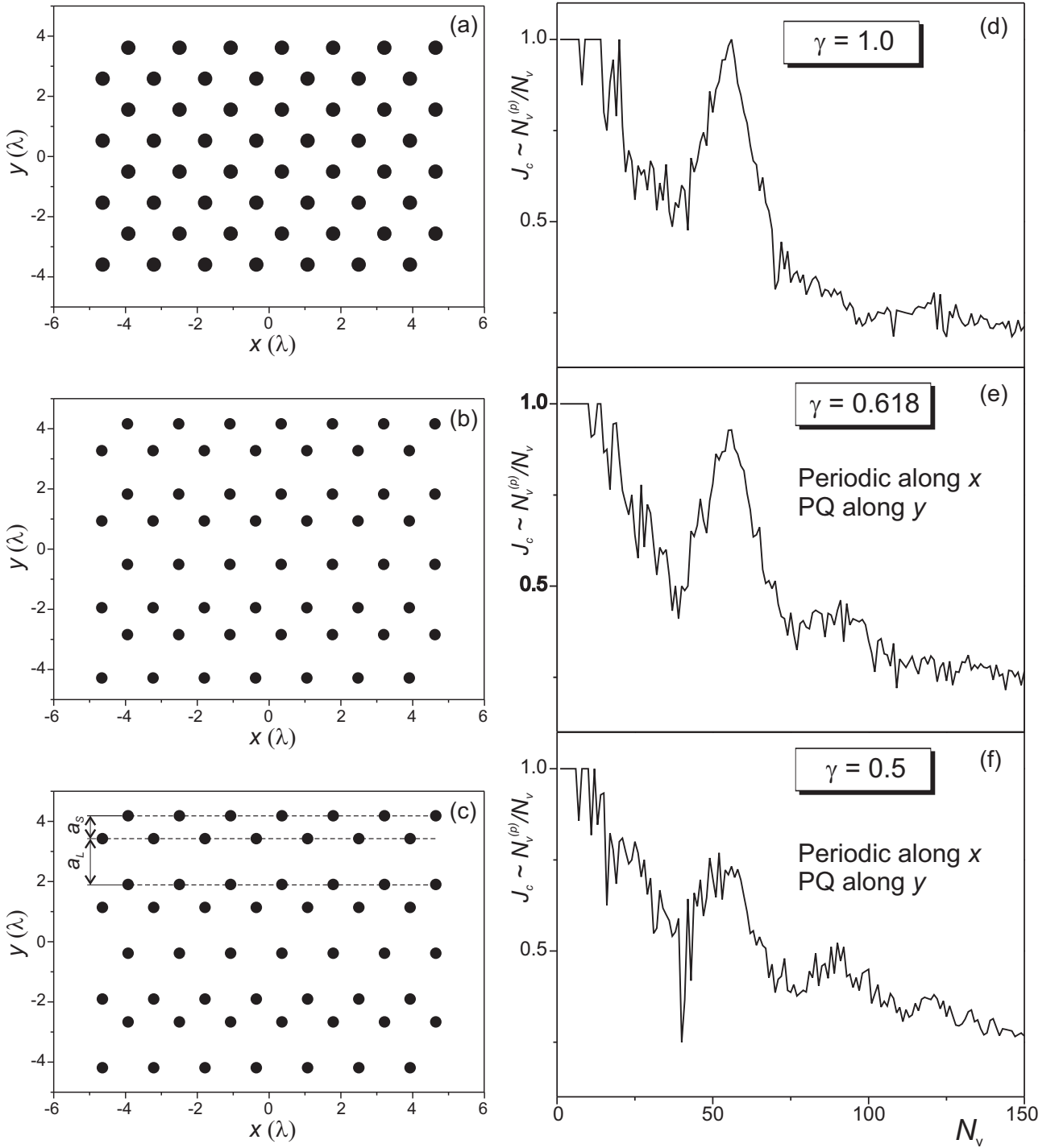}
\end{center}
\vspace{-5.0cm} 
\caption{ 
The spatial distribution of pinning sites for a 
triangular lattice (i.e., periodic) (a), 
and 2D-quasiperiodic (2DQP) triangular, i.e. periodic along 
one direction (the $x$-direction) and QP along
the other one (the $y$-direction), (b) and (c). 
The parameter 
$\gamma_{y}$ 
is defined as a ratio of short ($a_{S}$) to long ($a_{L}$) periods 
in the $y$-direction, as shown in (c). 
The values of the parameter 
$\gamma_{y}$ 
are:
$\gamma_{y} = 1.0$ (a), 
$\gamma_{y} = 1 / \tau,$ where $\tau = (1 + \sqrt{5})/2$ (golden mean) (b), 
$\gamma_{y} = 0.5$ (c).
The critical depinning current 
$J_{c}$, 
as a function of 
the number of vortices, 
$N_{v} \sim \Phi$, 
for triangular (d) and 
the 2DQP triangular 
lattices [shown in (b) and (c)] 
(e) and (f), correspondingly, 
for 
$f_{p}/f_{0} = 2.0$,
$r_{p} = 0.1\lambda$.
}
\end{figure}

\end{widetext}

\noindent 
in the other direction 
($y$-direction).


\subsection{
2D-quasiperiodic triangular lattice 
of pinning sites}

In an infinitely long one-dimensional homogeneous superconductor without any pinning centers, 
vortices obviously are equidistantly distributed, forming a periodic chain. 
Similarly, as it is well-known, 
in a three-dimensional superconductor (or in quasi-two-dimensional slabs or films), 
vortices organize themselves in a periodic triangular lattice shown schematically in Fig.~6a. 
If we keep the lattice undistorted along the 
$x$-direction
and introduce a 
{\it quasiperiodic deformation }
along the 
$y$-direction, 
similarly to the case of a 1D QP chain,
we obtain a 
2QP triangular lattice
as shown in Figs.~6b,c. 
The ``quasiperiodicity'' parameter 
$\gamma_{y}$
for this 2DQP lattice 
is defined as the ratio of the short to long periods, 
$a_{L}$ to $a_{S}$
(see Fig.~6c).
The 2DQP triangular arrays of pinning sites are shown for 
$\gamma_{y} = 1.0$ (Fig.~6a), 
$\gamma_{y} = 1 / \tau,$ where $\tau = (1 + \sqrt{5})/2$ (Fig.~6b), 
$\gamma_{y} = 0.5$ (Fig.~6c).
The corresponding functions 
$J_{c}(N_{v} \sim \Phi)$
are presented in Figs.~6d,e,f
for
the following pinning parameters: 
$f_{p}/f_{0} = 2.0$, 
and 
$r_{p} = 0.1\lambda$. 

For the triangular array of pinning sites (Fig.~6a), 
we obtain a well-resolved main commensurability peak (Fig.~6d),
corresponding to the first matching field, at
$\Phi = \Phi_{1}$.
Note that the vortex lattice is in general 
incommensurate with a triangular
lattice of pinning sites for the second matching field, i.e. when
$\Phi = 2\Phi_{1}$ 
(see, e.g., Ref.~\cite{md0157}).
For instance, 
for the parameters used in our simulations, 
only each second row of pinning sites is occupied at 
$\Phi = 2\Phi_{1}$, 
resulting in a very weak maximum of the function
$J_{c}(N_{v} \sim \Phi)$
at that point;
the parameters used 
($f_{p}/f_{0} = 2.0$,
$r_{p} = 0.1\lambda$)
are nearly optimal for revealing features of the function 
$J_{c}(\Phi)$
related to quasiperiodicity. 

When tuning 
$\gamma$
out of the periodic value 
$\gamma_{y} = 1.0$, 
the main peak decreases in magnitude, 
and a maximum forms near it 
at a larger value of 
$N_{v} \sim \Phi$. 
These changes are demonstrated in Fig.~6e 
(the corresponding pinning array is shown in Fig.~6b)
for 
$\gamma_{y} = 1 / \tau$. 

It should be noted that 
here the 
parameter 
$\gamma_{y}$ 
has a different meaning,
in the case of triangular 2DQP
lattices, 
compared to the case of 
$\gamma$ 
for the 
1D QP chains
considered in the previous section. 
In a 1D QP chain, 
$\gamma = a_{S}/a_{L}$ 
is the ratio of distances between pinning sites, 
whereas 
in a triangular 2DQP lattice 
$\gamma_{y}$ 
defines the ratio of the distances between the rows of pinning sites
(see Fig.~6c). 
It is easy to show that the ratio of distances between the neighboring pinning
sites in a triangular 2DQP lattice is
\begin{equation}
\gamma^{\prime} = \sqrt{ \frac{ (1 + \gamma_{y}^{2})a^{2} + 12 a^{2} }{ (1 + \gamma_{y}^{2})a^{2} 
+ 12 \gamma_{y}^{2} a^{2} } },
\label{gammaqqp}
\end{equation}
where
$a$
is the period of the 2DQP triangular lattice along the (periodic)
$x$-direction. 
Thus, 
for 
$\gamma_{y} = 1 / \tau$, 
the parameter 
$\gamma^{\prime}$
defined by Eq.~(\ref{gammaqqp})
becomes 
$\gamma^{\prime} \approx 0.7$. 

For 
$\gamma_{y} = 0.5$
($\gamma^{\prime} \approx 0.6$), 
the function 
$J_{c}(N_{v})$
is plotted in Fig.~6f. 
The main peak is further depressed, 
while the closest satellite peak becomes more pronounced. 
In addition, other satellite peaks appear, which are much less pronounced. 

\vspace{1.0cm}

\section{Transition from a triangular to a quasiperiodic Penrose-lattice 
array of pinning sites } 


Above, we have revealed some features of the 
behavior of the critical depinning current 
$J_{c}$
as a function of the applied magnetic flux 
$\Phi$
(or as a function of the number of vortices in the system
$N_{v} \sim \Phi$). 
They have been found under the transformation of a triangular lattice 
to a 2DQP triangular lattice (array) of pinning sites. 

Consider now a similar procedure 
but the final configuration of the transformation 
from a triangular lattice
will be a 
2D QP array of pinning sites, namely, 
an array of pinning sites located at the nodes of a 
{\it five-fold Penrose lattice. }
This kind of lattice is representative of a class of 2D QP
structures, or quasicrystals, which are referred to as Penrose tilings. 
These structures possess a local order 
and 
{\it a rotational} 
(five- or ten-fold) symmetry, 
but do {\it not} have {\it translational long-range order.}
Being constructed of a series of building blocks of certain simple shapes
combined according to specific local rules, 
these structures can extend to infinity without any defects \cite{bookqc}. 
Below, we will discuss in more detail the structure of the Penrose lattice. 

The transformation of a triangular lattice to the Penrose one is 
a rather non-trivial procedure,
as distinct from the transformation to a 2DQP triangular lattice
done above
when we simply 
stretched 
some of the inter-row distances and 
squeezed 
other ones according to the Fibonacci rules (Eq.~(\ref{rule}))
for a one-dimensional QP lattice. 
In order to find intermediate configurations between the triangular
lattice and the Penrose lattice, we employ the following approach. 
First, we place non-interacting vortices
at the positions coinciding with the nodes of the Penrose lattice
(these can be considered as pinning sites for vortices, which 
right afterwards are ``switched off''). 
Then we let the vortices freely relax undergoing the vortex-vortex interaction 
force at low temperatures (and no pinning force). 
The vortices relax to their ground state, which is a triangular lattice. 
During the relaxation process, we do a series of ``snapshots'', 
recording the coordinates of the vortex configurations at different times. 
The sets of coordinates obtained are then used as coordinates of pinning sites. 
We arrange these sets in antichronological order to model
a continuous transition of the pinning array from its initial configuration, 
a triangular lattice, to its final configuration, a Penrose


\begin{figure}[btp]
\begin{center}
\end{center}
\caption{ 
(This figure is available in ``png'' format: ``Penrose Fig 7.png''; color)
Left column: 
Transformation of a triangular lattice of pinning sites to 
a five-fold Penrose lattice.
The distributions of the pinning sites in the triangular lattice 
(shown by blue solid circles in (a) and by blue open circles in (b), (c), and (d),
for comparison). 
Intermediate configurations (shown by red solid circles in (b) and (c)) between 
the triangular (a) and the Penrose lattice 
(shown by red solid circles in (d)). 
Right column: 
The corresponding critical depinning currents, 
$J_{c}$, 
as a function of 
the applied magnetic flux, 
$\Phi$, 
for the pinning arrays shown in (a) to (d), respectively: 
for the triangular lattice
(shown by blue solid line in (e) and by blue dashed lines in (f), (g), and (h), 
for comparison);
for the intermediate configurations 
(shown by red solid lines in (f) and (g)); 
for the Penrose lattice 
(shown by the red solid line in (h)). 
For the Penrose lattice case in (b), 
the drop in $J_{c}(\Phi)$ is an artifact of the boundary conditions. 
Namely, the Penrose lattices of pinning sites did not fit the square cell 
used in the simulations. 
Thus, the freely-moving vortices near the edges significantly decreased 
the value of $J_{c}$, especially near $\Phi_{1}$. 
This problem will be dealt separately in Fig.~14 and in Eq.~(\ref{eta}). 
}
\end{figure}


\noindent
lattice. 

In Figs.~7a to 7d, 
four of these configurations of pinning sites are shown. 
The triangular lattice is presented in Fig.~7a. 
Two intermediate configurations are shown in Figs.~7b,c. 
The pinning sites plotted in Fig.~7d 
are located on the vertices of a Penrose lattice. 
For comparison, 
a pinning array in the form of a triangular lattice (Fig.~7a) 
is also presented in Figs.~7b,c,d 
as open blue circles. 
The functions 
$J_{c}(\Phi)$
calculated for the pinning arrays shown in 
Figs.~7a to d, 
are plotted, respectively, in
Figs.~7e to h. 
The function 
$J_{c}(\Phi)$
for the triangular lattice (Fig.~7e)
is also plotted, for comparison, in 
Figs.~7g,f,h 
as a blue dashed curve. 

The main commensurability peak related to the first matching field 
in a triangular lattice of pinning sites,
observed at 
$\Phi = \Phi_{1}$ 
(Fig.~7e),
turns out to be rather stable with respect to moderate deformations
of the lattice
(Fig.~7f, see also Fig.~7b). 
It still has a maximum height in Fig.~7f, 
although it broadens.
However, 
the depths of the valleys near the peak decreases by about 20 to 30 per cent. 
Two 
sharp peaks near
$\Phi = \Phi_{1}/3$ 
and
$\Phi = \Phi_{1}/6$ 
(Fig.~7e),
related to the commensurability of the long-range order 
in a triangular lattice,
disappear. 
With further deformation, 
e.g., 
for the pinning array shown in Fig.~7c, 
the main peak still remains
but only about 80 per cent
of the vortices are pinned in this case. 
The function 
$J_{c}(\Phi)$
becomes somewhat smoother,
and it does not display any pronounced features 
(for $\Phi \gtrsim 2\Phi_{1}/3$) 
except the main maximum. 

The transition to 
a Penrose-lattice array of pinning sites 
(Fig.~7d)
is accompanied by 
the appearance of a specific fine structure
of the function
$J_{c}(\Phi)$.
Namely, 
two well-resolved features on the broad main maximum
(Fig.~7h)
are the most pronounced ones. 
Other, less pronounced, features will be discussed 
below 
for larger Penrose-lattice arrays.
For large arrays, 
the function
$J_{c}(\Phi)$
is much less affected by 
fluctuations 
related to the entrance of each single vortex in the system, 
which are significant for the small-size array shown in Fig.~7d 
($N_{p} = 56$).
This small-size array is 
used here just as an illustration, 
for studying the transition form a periodic (triangular) 
to a QP (Penrose lattice) pinning site array. 
However, 
studying even 
a relatively small piece of a QP structure provides 
some useful information about properties of the whole 
system based on the self-similarity of the lattice, 
which was revealed for 1D QP chains 
in the previous section, 
and which will be demonstrated below for 
2D QP structures. 

The Penrose and the 2DQP triangular lattice 
(Fig.~6f) both have an important similar feature: 
their $J_{c}(\Phi)$ has two nearby maxima. 
Thus, 
our previous analysis based on 
several alternative ways to continuously deform a periodic lattice to a QP one 
shows that the features shown are 
hallmarks 
of QP pinning arrays. 

In the next section, 
the origin of the features observed will be explained
on the basis of
a detailed analysis 
of the structure and 
of the building blocks forming a five-fold Penrose lattice. 
Other, less pronounced, features will also be discussed. 
Some of them will be found in larger arrays 
in our numerical simulations.

\section{Analysis of the fine structures of the function 
$J_{c}(\Phi)$
in a quasiperiodic Penrose-lattice array of pinning sites}


\begin{figure}[btp]
\begin{center}
\vspace*{-0.5cm} 
\hspace{-1.0cm}
\includegraphics*[width=9.5cm]{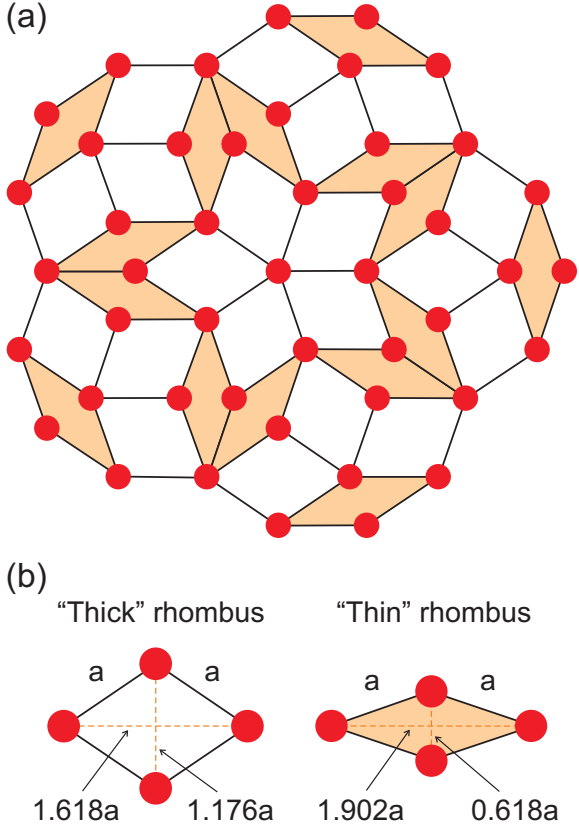}
\end{center}
\vspace{-1.5cm} 
\caption{ 
(Color)
The structure of a five-fold Penrose lattice (a). 
The elemental building blocks are rhombuses 
with equal sides 
$a$
and angles which are multiples of 
$\theta = 36^{o}$.
There are rhombuses of two kinds:
those having angles 
$2 \theta$
and 
$3 \theta$
(so called ``thick''), 
and 
rhombuses with angles
$\theta$
and 
$4 \theta$
(so called ``thin'') (b). 
The distances between the nodes,
i.e.
the lengths of the diagonals of the rhombuses
are: 
$1.176a$ (the short diagonal of a thick rhombus);
$(1 + \sqrt{5})a/2 = \tau a \approx 1.618a$
(the long diagonal of a thick rhombus);
$[(1 + \sqrt{5})/2 -1]a = (\tau -1)a  \approx 0.618a$, 
(the short diagonal of a thin rhombus);
$1.902a$ (the short diagonal of a thin rhombus) (b).
}
\end{figure}

The structure of a five-fold Penrose lattice is shown in Fig.~8.
As an illustration, a five-fold symmetric fragment which consists of 
$46$ points (nodes of the lattice) is presented (Fig.~8a). 
According to specific rules, 
the points are connected by lines
in order to display the structure of the Penrose lattice 
(compare, e.g., to Fig.~7d). 
The elemental building blocks are rhombuses 
with equal sides 
$a$
snd angles which are multiples of 
$\theta = 36^{o}$.
There are rhombuses of 
two kinds 
forming the Penrose lattice (Fig.~8b):
(i) those having angles 
$2 \theta$
and 
$3 \theta$
(so called ``thick''; they are empty in Fig.~8), 
and 
(ii) rhombuses with angles
$\theta$
and 
$4 \theta$
(so called ``thin''; they are colored in orange in Fig.~8). 

Let us analyze 
the structure of the Penrose lattice from the point of view
ot its pinning properties, 
when pinning sites are placed in the vertices of the lattice. 
In particular, 
we are interested whether any specific matching effects 
can exist in this system 
between the pinning lattice and the interacting vortices, 
which define the critical depinning current at 
different values of the applied magnetic field 
(i.e., the function $J_{c}(\Phi)$). 

On the one hand, 
QP (quasicrystalline) patterns are intrinsically 
{\it incommensurate} 
with the flux lattice for 
{\it any} 
value of the magnetic field 
\cite{behrooz,fnniu87,fnniu88,niufn89}, 
therefore, 
in contrast to periodic (e.g., triangular or square) pinning arrays,
one might 
{\it a priori} 
assume a {\it lack} of sharp peaks in 
$J_{c}(\Phi)$ 
for QP arrays of pinning sites. 

On the other hand, 
the existence of many periods in the Penrose lattice 
can lead to a hierarchy of 
matching effects for certain values of the applied magnetic field, 
resulting in strikingly-broad shapes for $J_{c}(\Phi)$. 

In order to match the vortex lattice on an entire QP pinning array, 
the specific geometry of the elements which form the QP lattice is important 
as well as their arrangement, 
as distinct from
the flux quantization effects
and superconductor-to-normal phase boundaries
for which the areas of the elements only plays a role
\cite{fnniu87,fnniu88,niufn89}. 
As mentioned above, 
a five-fold Penrose lattice is constructed of building blocks, 
or rhombuses, of two kinds. 
While the sides of the rhombuses are equal
(denoted by $a$), 
the distances between the nodes
(where we place pinning sites)
are 
{\it not} 
equal (which is problematic for vortices). 
The lengths of the diagonals of the rhombuses 
are as follows (Fig.~8b): 
$1.176a$ (the short diagonal of a thick rhombus);
$(1 + \sqrt{5})a/2 = \tau a \approx 1.618a$, 
where 
$\tau$
is the golden mean
(the long diagonal of a thick rhombus);
$[(1 + \sqrt{5})/2 -1]a = (\tau -1)a \approx 0.618a$, 
(the short diagonal of a thin rhombus);
$1.902a$ (the short diagonal of a thin rhombus).

Based on this hierarchy of distances, 
we can predict matching effects 
(and corresponding features of the function 
$J_{c}(\Phi)$)
for the Penrose-lattice pinning array. 

First, we can expect that there is a ``first matching field'' 
(let us denote the corresponding flux as $\Phi_{1}$) 
when each pinning site is occupied by a vortex. 
Although sides of all the rhombuses are equal 
to each other 
similarly to that in a periodic lattice, 
nevertheless 
this matching effect 
is not expected to be 
accompanied by a sharp peak. 
Instead, 
it is a 
{\it broad }
maximum
since it involves 
{\it three }
kinds of local 
``commensurability'' effects of the flux lattice:
with the rhombus side 
$a$; 
with the short diagonal of a thick rhombus, 
$1.176a$, 
which is close to 
$a$; 
and 
with the short diagonal of a thin rhombus, 
which is the golden mean times
$a$, 
$0.618a$ (see Fig.~8b). 

It should be noted that 
this kind of matching 
assumes 
that a vortex lattice is rather 
{\it weak}, 
i.e.
the effect can be more or less pronounced depending on 
the specific relations between 
the vortex-vortex interaction constant and 
the strength of the pinning sites 
as well as 
on the distance between pinning sites
and their radius. 
Assuming that 
the vortex-vortex interaction constant is a material parameter, 
all others 
can be 
adjustable parameters in experiments with
artificially-created QP pinning arrays. 
For instance, 
the pinning parameters can be 
``adjusted''
by using as pinning centers 
antidots, 
i.e.
microholes of different radii ``drilled'' in a superconductor film 
\cite{vvmdotprl,vvmdotprb,rwdot}, 
or 
blind antidots 
\cite{vvmbdot} 
of different depths and radii. 

Further, 
we can deduce 
that next to the above ``main'' matching flux
there is another matching related with
the filling of all the pinning sites in the vertices of thick rhombuses
and only 
{\it three out of four}
of the pinning sites in the vertices of thin rhombuses,
i.e., one of the pinning sites 
in the vertices of thin rhombuses is empty. 
For this value of the flux, 
which is lower than 
$\Phi_{1}$, 
matching conditions are fulfilled for 
{\it two }
close distances, 
$a$ (the side of a rhombus)
and 
$1.176a$ (the short diagonal of a thick rhombus)
but 
are not fulfilled for 
the short diagonal, 
$a/\tau$, 
of the thin rhombus. 

Therefore, 
this 
QP feature 
is related to 
the golden mean value, 
although not in such a direct way as in the case of a 1D QP pinning array. 
This 2D QP matching results in a very wide maximum of the function 
$J_{c}(\Phi)$.
The position of this broad maximum, 
i.e., 
the specific value of 
$\Phi$
(denoted here by 
$\Phi_{\rm vacancy/thin} \equiv \Phi_{\rm v/t} = 0.757 \, \Phi_{1}$) 
could be found as follows. 
The ratio of the numbers of thick and thin rhombuses is determined by the 
Fibonacci numbers 
and in the limit of large pinning arrays,
$N_{p} \to \infty$
this ratio 
tends to the golden mean. 
The number of unoccupied pinning sites is governed by the number of thin rhombuses. 
However, some of the thin rhombuses are separated from other thin rhombuses by thick ones
(call them 
{\it single} 
thin rhombuses), 
but some of them have common sides with each other 
({\it double} 
thin rhombuses). 
Therefore, the number of vacancies (i.e., unoccupied pins) is then the number of single thin rhombuses 
plus one 
{\it half }
of the number of ``double'' thin rhombuses, 
\begin{equation}
N_{p}^{\rm un} (\Phi_{\rm v/t}) = N_{\rm rh}^{\rm s} + \frac{1}{2} N_{\rm rh}^{\rm d},
\label{ssds}
\end{equation}
where
$N_{p}^{\rm un}$ 
is the number of unoccupied pinning sites at 
$\Phi = \Phi_{\rm v/t}$, 
$N_{\rm rh}^{\rm s}$
and
$N_{\rm rh}^{\rm d}$
are the numbers of single and double thin rhombuses, correspondingly. 

For higher vortex densities
(e.g., for
$\Phi = \Phi_{\rm interstitial/thick} \equiv \Phi_{\rm i/T} = 1.482 \, \Phi_{1}$), 
we can expect the appearance of a feature 
(maximum)
of the function 
$J_{c}(\Phi)$
related to the 
entry of a single {\it interstitial} vortex into each {\it thick} rhombus. 
The position of this maximum 
is determined by the number of vortices at 
$\Phi = \Phi_{1}$,
which is 
$N_{v}(\Phi) = N_{p}$,
plus the number of thick rhombuses, 
$N_{\rm rh}^{\rm thick}$. 
Since the ratio of the number of thick to that of thin rhombuses
is the golden mean, 
$\tau = (1 + \sqrt{5})/2$
(in an infinite lattice; in a finite lattice,


\begin{figure}[btp]
\begin{center}
\end{center}
\caption{ 
(This figure is available in ``png'' format: ``Penrose Fig 9.png''; color)
The critical depinning current 
$J_{c}$
as a function of 
the applied magnetic flux, 
$\Phi \sim N_{v}$, 
for an array of pinning sites placed 
at the nodes of a five-fold Penrose lattice 
(for a part of the lattice which contains 
$N_{p} = 46$ 
pinning sites) (a). 
The distributions of vortices 
(shown by green dots) 
pinned on 
the Penrose-lattice pinning site array
(pinning sites are shown by red open circles 
connected by orange solid lines used in order to
show the Penrose lattice structure, 
i.e., thick and thin rhombuses), 
for specific values of the applied magnetic flux:
(b) 
$\Phi = \Phi_{\rm vacancy/thin} \equiv \Phi_{\rm v/t} = 0.757 \, \Phi_{1}$, 
vortices occupy all the pinning sites
except those in 
one of the two vertices (connected by the short diagonal) of each thin rhombus,
each single and each 
pair of double
thin rhombuses contain one unoccupied pinning site
at the matching field
$\Phi_{\rm v/t}$; 
(c)
$\Phi = \Phi_{1}$, 
the number of vortices 
$N_{v}$
coincides with the number of pinning sites
$N_{p}$, 
and almost all the vortices are pinned 
(because of using a square simulation cell, some of the vortices 
are always ``interstitial''
but allow to keep the average vortex density
in the entire simulation cell;
due to these additional vortices, 
the value of the function 
$J_{c}(\Phi)$ 
effectively reduces 
by a ``filling'' factor 
$\eta = A_{P}/A \approx 0.575$, 
where 
$A_{P}$ 
and 
$A$ 
are the areas 
of the Penrose-lattice ``sample''
and 
of the simulation region; 
(d)
$\Phi = \Phi_{\rm interstitial/thick} \equiv \Phi_{\rm i/T} = 1.482 \, \Phi_{1}$, 
vortices occupy all
the pinning sites 
and interstitial positions inside each thick rhombus, 
one vortex per each thick rhombus.
The parameters are
$f_{p}/f_{0} = 2.0$,
$r_{p} = 0.1\lambda$.
}
\end{figure}


\noindent 
it is determined by a ratio of 
two successive Fibonacci numbers),
then
$N_{\rm rh}^{\rm thick} = N_{\rm rh}/\tau$, 
where 
$N_{\rm rh}$
is the total number of rhombuses. 
Here we used:
$1/\tau = \tau - 1.$

In Fig.~9a, 
the function 
$J_{c}(\Phi)$ 
is plotted 
for an array of pinning sites in the form of a part of the Penrose lattice,
shown in Figs.~9c,d,e, 
which consists of 20 thick 
($N_{\rm rh}^{\rm thick} = 20$) 
and 15 thin 
($N_{\rm rh}^{\rm thin} = 15$) 
rhombuses
and contains
46 nodes
(pinning sites). 
The nodes are connected by lines in order to 
show the rhombuses. 

The distribution of vortices for 
$\Phi = \Phi_{1}$
is shown in 
Fig.~9c. 
The number of vortices 
$N_{v}$
coincides with the number of pinning sites
$N_{p}$, 
and almost all the vortices are pinned.
Note that 
since we use a square simulation cell, some of the
vortices 
are always outside the ``Penrose sample''. 
These vortices 
mimic the externally applied magnetic field and determine 
the average vortex density
in the entire simulation cell. 
Because of these additional vortices, 
the value of the function 
$J_{c}(\Phi)$ 
effectively reduces 
approximately by a ``filling'' factor 
$\eta$
which is 
\begin{equation}
\eta = \frac{A_{P}}{A} \approx 0.575. 
\label{eta}
\end{equation}
Here 
$A_{P}$ 
and 
$A$ 
are the areas 
of the Penrose lattice (i.e., the area of all the rhombuses)
and 
of the simulation region. 

The value of the function 
$J_{c}(\Phi)$
in the maximum 
$\Phi = \Phi_{1}$
(Fig.~9a)
is 
$J_{c} \approx 0.55$, 
i.e.
corresponds to almost 
perfect matching 
(two pinning sites occurred to be unoccupied in the distribution
shown in Fig.~9c)
taking into account Eq.~(\ref{eta}).

Let us now more carefully analyze 
the calculations of $J_{c}(\Phi)$ for the Penrose lattice, presented in Fig.~9. 
In Fig.~9b, 
the distribution of vortices is shown for
$\Phi = \Phi_{\rm v/t}$.
Vortices occupy all the pinning sites except those situated in one 
of the two vertices, connected by the short diagonal, of each thin rhombus. 
Thus, each single and each pair of double
thin rhombuses contain one vacancy (unoccupied pinning site) 
at the matching field 
$\Phi_{\rm v/t}$.
The corresponding maximum is indicated by the arrow (b) 
in Fig.~9a. 

The location of vortices for 
$\Phi = \Phi_{1}$
(the maximum (c) in Fig.~9a) 
is shown in 
Fig.~9c; 
here the number of vortices 
$N_{v}$
coincides with the number of pinning sites
$N_{p}$, 
and almost all the vortices are pinned.

The distribution of vortices for 
$\Phi = \Phi_{\rm i/T}$
(Fig.~9d)
is also in agreement with our expectation: 
vortices occupy all the pinning sites 
(there is only a single ``defect'' in the distribution shown in Fig.~9d:
one vortex left the pinning site and became interstitial)
plus interstitial positions inside each thick rhombus, 
i.e., one vortex per each thick rhombus. 
However, 
the corresponding feature 
of the function 
$J_{c}(\Phi)$
(arrow (d) in Fig.~9a) 
is less pronounced 
than the two above maxima at 
$\Phi = \Phi_{1}$
and at
$\Phi = \Phi_{\rm v/t}$.

In addition, 
there is a weak feature 
of the function 
$J_{c}(\Phi)$
at 
$\Phi \approx \Phi_{\rm v/t}/2$, 
which more clearly manifests itself 
for larger Penrose-lattice pinning arrays
(see Fig.~10a). 

Therefore, 
the calculated distributions of the vortices 
pinned on the Penrose-lattice pinning site array
and the resulting 
function 
$J_{c}(\Phi)$
have revealed the QP features 
which are in agreement with our expectations. 
The specific structure of 
the function
$J_{c}(\Phi)$ 
is consistent with two previous derivations both based on 
continuously deforming a QP lattice into a Penrose one (Sections V and VI).

\begin{figure}[btp]
\begin{center}
\hspace*{-4.0cm}
\end{center}
\caption{ 
(This figure is available in ``png'' format: ``Penrose Fig 10.png''; color)
The critical depinning current 
$J_{c}$
as a function of 
the applied magnetic flux, 
$\Phi \sim N_{v}$, 
for an array of pinning sites placed 
in the nodes of a five-fold Penrose lattice
(for $N_{p} = 301$ ) (a). 
The distributions of vortices 
(shown by green solid circles)
pinned on 
the Penrose-lattice pinning site array
(shown by red open circles), 
for specific values of the applied magnetic flux
which correspond to two matching fields:
(b) $\Phi = \Phi_{\rm v/t}$, 
vortices occupy all the pinning sites
except one in each thin rhombus; 
(c) $\Phi = \Phi_{1}$, 
all the vortices are pinned. 
The parameters are 
the same as in Fig.~9.
}
\end{figure}

In Fig.~10a, 
the function 
$J_{c}(N_{v})$
is shown 
for a larger Penrose-lattice array of pinning sites, 
$N_{p} = 301$. 
The above QP features in
$J_{c}(N_{v})$
are much more pronounced
in this case then for smaller arrays
because of a considerable reduction of 
the ``noise'' related with an entry of 
each single vortex in the system. 

In particular, 
the main maximum
of the function 
$J_{c}(N_{v})$,
which corresponds to the matching condition 
$\Phi = \Phi_{1}$
($N_{v} = 301$),
transforms into a rather sharp 
peak with the magnitude 
$\eta$. 
Also, a local maximum of 
$J_{c}(\Phi)$
at 
$\Phi \approx \Phi_{\rm v/t}/2$, 
is more pronounced
for 
$N_{p} = 301$, 
as mentioned above.

\begin{figure}[btp]
\begin{center}
\hspace{-1.0cm}
\includegraphics*[width=9.5cm]{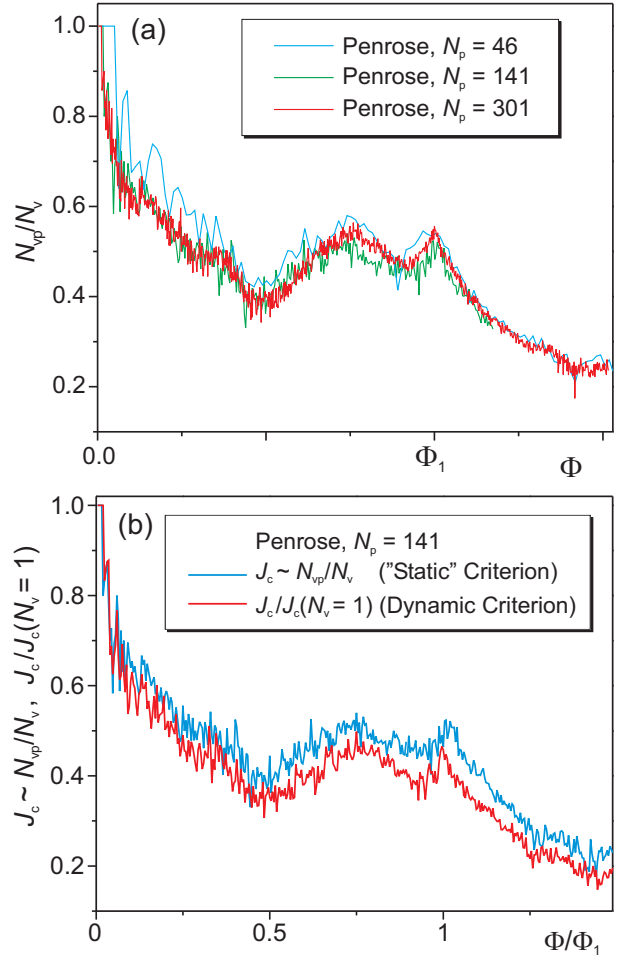}
\end{center}
\vspace{-1.0cm} 
\caption{ 
(Color)
(a) The critical depinning current 
$J_{c}$
as a function of 
the applied magnetic flux, 
$\Phi \sim N_{v}^{(p)}/N_{v}$, 
shown for three different
Penrose-lattice pinning site arrays:
$N_{p} = 46$
(shown by blue solid line),
$N_{p} = 141$
(shown by green solid line),
$N_{p} = 301$
(shown by red solid line). 
The revealed features are hallmarks of
a Penrose-lattice pinning site array. 
(b) Comparison of $J_{c}$ 
as a function of 
the applied magnetic flux, 
calculated using the ``static'' 
and dynamical criteria. 
The latter means calculating $J_{c}$ using a threshold criterion, 
i.e., $J_{c}$ is obtained as the minimum current $J \  \propto \; f_{\rm i}^{d}$ 
which depins the vortices. 
The results obtained using these two criteria are essentially equal.
}
\end{figure}

Finally, 
Fig.~11
demonstrates 
the function 
$J_{c}(\Phi)$, 
calculated for different samples 
with 
$N_{p} = 46$, 141, and 301 (Fig.~11a), 
and also for different criteria of 
$J_{c}$: 
for the ``static'' and dynamical criteria (Fig.~11b). 
In the dynamical simulations of $J_{c}$ using a threshold criterion, 
i.e., $J_{c}$ is obtained as the minimum current $J \  \propto \; f_{\rm i}^{d}$ 
which depins the vortices. 
In the Appendix A, we show onset of vortex motion when the applied current $J$ exceeds the 
critical currnet: $J > J_{c}$. 
The results obtained using these two criteria are essentially equal, and throughout this 
work we use the ``static'' criterion defined above.

\section{Analytical approach}


The following analysis could provide a better understanding 
of the above 
structure of $J_c(\Phi)$ for the Penrose pinning 
lattice. 
Let us compare the elastic $E_{\rm el}$ and 
pinning $E_{\rm pin}$ energies of the vortex lattice at 
$H_1$ and at (the lower field) $H_{\rm v/t}$, 
corresponding to the two maxima of $J_c$ (e.g., Figs.~9a, 10a, 11). 
Vortices can be pinned 
if the gain $E_{\rm pin}= U_{\rm pin} \, \beta \, n_{\rm pin}$
of the pinning energy is larger than the increase of the elastic
energy 
\cite{brandt,vinokur,clem}
related to local compressions: 
\begin{equation}
E_{\rm el}=C_{11} \frac{(a_{\rm eq}-b)^2}{a_{\rm eq}}. 
\label{eel} 
\end{equation}
The shear elastic energy ($\propto C_{66}$) provides the same qualitative result. 
%
Here, $U_{\rm pin}\sim f_p \, r_p$, 
$n_{\rm pin}$ is the density of pinning centers, 
%
$$\beta (H \leq H_{1}) = H/H_{1} = B/(\Phi_{0} n_{\rm pin}),$$ 
and $\beta (H > H_{1}) = 1$ 
is the fraction of occupied pinning sites
($\beta=1$ for $H=H_1$, and $\beta=0.757$ for $H=H_{\rm v/t}$), 
$a_{\rm eq}=\left( 2/\sqrt{3}\beta n_{\rm pin} \right)^{1/2}$ is the equilibrium 
distance between vortices in the triangular lattice, 
$b$ is the minimum distance between
vortices in the distorted pinned vortex lattice ($b=a/\tau$ for
$H=H_1$ and $b=a$ for $H=H_{\rm v/t}$), and 
\begin{equation}
C_{11}= \frac{B^2}{4 \pi (1+\lambda^2 k^2)} 
\label{c11} 
\end{equation}
is the compressibility modulus 
for short-range deformations 
\cite{brandt}
with characteristic spatial scale 
$k \approx \left( n_{\rm pin} \right)^{1/2}$. 
The dimensionless difference of
the pinning and elastic energies is 
\begin{equation}
E_{\rm pin}-E_{\rm el} = \frac{ \beta f_{\rm diff} \, n_{\rm pin} \Phi_0^2 }{4\pi\lambda^2}, 
\label{epin} 
\end{equation}
where
\begin{equation}
f_{\rm diff}=\frac{4\pi\lambda^2 U_{\rm pin}}{\Phi_0^2}
-\beta\left[1 - b \left( \frac{\beta \sqrt{3} n_{\rm pin} }{2} \right)^{1/2} \right]^2.
\label{fdiff}
\end{equation}
The function 
$f_{\rm diff}0$ 
is shown schematically in the inset to Fig.~12. 
Near matching fields, $J_c$ has a peak when $f_{\rm diff} > 0$
(and no peak when $f_{\rm diff} < 0$). 
Since only two matching fields provide $f_{\rm diff} > 0$, 
then our analysis explains the two-peak structure observed in $J_c$ 
shown in Figs.~9a, 10a, 11. 
For instance, for the main matching fields 
Eq.~(\ref{fdiff}) gives: 
$f_{\rm diff}(\Phi_{\rm v/t}) \approx 0.0056$, 
$f_{\rm diff}(\Phi_{1}) \approx 0.0058$, 
and 
$f_{\rm diff}(\Phi_{\rm i/T}) \approx -0.09$. 
Note that for weaker pinning, the two-peak structure gradually turns into 
one very broad peak, and eventually zero peaks for weak enough pinning (see Fig.~12). 
The $J_{c}$ peaks corresponding to higher
matching fields are strongly suppressed because of the fast increase
($\propto B^2$) of the compressibility modulus $C_{11}$ and, thus, the elastic
energy with respect to the pinning energy; 
the latter cannot exceed the maximum value $U_{\rm pin}n_{\rm pin}$. 
The subharmonic peaks of
$J_c$, which could occur for lower fields $H < H_{\rm v/t}$, are also 
suppressed due to the increase of $C_{11}$ associated with the growing 
spatial scales $1/k$ of the deformations.


\begin{figure}[btp]
\begin{center}
\vspace*{-6.0cm}
\hspace*{-0.5cm}
\includegraphics*[width=9.5cm]{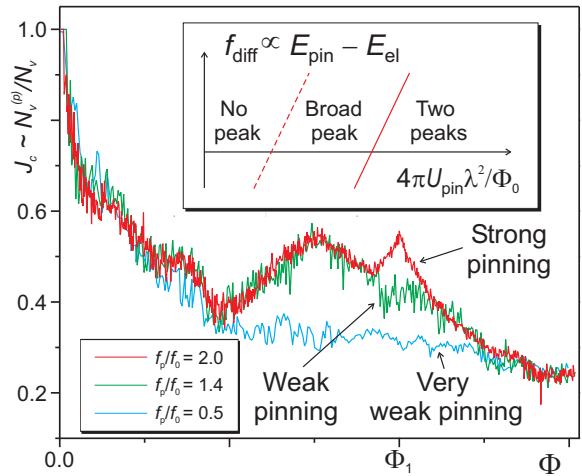}
\end{center}
\vspace{-1.0cm}
\caption{
(Color)
The critical current 
$J_{c}(\Phi)$ 
for Penrose-lattice arrays 
for different 
pinning strength 
$f_{p}/f_{0} = 2.0$ (red solid line),  
$f_{p}/f_{0} = 1.4$ (green solid line),  
$f_{p}/f_{0} = 0.5$ (blue solid line). 
The peak at $\Phi_{1}$ is suppressed for weaker pinning ($f_{p}/f_{0} = 1.4$). 
Eventually, all the main peaks disappear for sufficiently weak pins ($f_{p}/f_{0} = 0.5$). 
The inset shows the dimensionless difference, $f_{\rm diff}$, of the pinning and the elastic 
energies versus the pinning-to-interaction energy ratio, for the broad $J_{c}$ peak 
at $\Phi_{\rm v/t}$ (red dashed line) and for $\Phi_{1}$ (red solid line). 
Only $f_{\rm diff} > 0$ gives stable peaks in $J_{c}$. 
}
\end{figure}

\section{The critical current $J_{c}(\Phi)$
in a randomly distorted triangular lattice}


Above we have studied 
the function 
$J_{c}(\Phi)$
for periodic, QP 2D arrays of pinning sites
and analyzed the transition from 
the periodic triangular lattice to the QP Penrose lattice (see Fig.~7). 
One of the issues,  
which is 
related to this analysis and can be useful for practical applications, 
is the increase of the critical current 
(shown, e.g., in Fig.~7f)
in the regions
corresponding to 
{\it minima }
of 
$J_{c}(\Phi)$
for periodic (triangular) pinning arrays. 
The situation shown in Fig.~7f
seems to be the optimal from the point of view 
of a homogeneous increase of 
$J_{c}(\Phi)$
without degradation of the main peak at 
$\Phi = \Phi_{1}$. 
Recall that 
it corresponds to 
a 
{\it slightly }
``quasiperiodically distorted''
triangular lattice (see Fig.~7b). 
The pinning sites of the triangular lattice are shifted from their 
``correct'' positions
but not randomly:
their positions are determined 
by vortex-vortex interaction, 
which tries to restore the triangular lattice, 
and 
by the memory about previous configurations
including the initial one, 
i.e., 
the Penrose lattice. 
Analyzing the lattice presented in Fig.~7b
we can deduce that 
it keeps some short-range order
(i.e., distorted triangular cells similar 
in shape and size 
to those
in the triangular lattice)
but does not have long-range order of the triangular
lattice. 
As a result, the main peak remains, 
since it is related to the short-order matching effects
(i.e., over the distances of the order inter-site spacings $a$). 
The sharp decrease around the maximum and appearance of 
the deep valleys is explained by the absence 
(due to long-range order, i.e., over distances longer than $a$) 
of any matching effects 
for the flux densities close to 
$\Phi = \Phi_{1}$ 
(there is no matching for 
$\Phi = \Phi_{1}/2$ 
or 
$\Phi = 1.5\Phi_{1}$ 
and 
$\Phi = 2\Phi_{1}$ 
for the triangular lattice). 
When the long-range order is destroyed,
as shown in Fig.~7b, 
matching effects other than 
$\Phi = \Phi_{1}$
become allowed. 

It is appropriate to mention here that 
the QP Penrose lattice possesses 
a short-range order but
does not have
a long-range translational order. 
In such a way, 
the 
``quasiperiodically distorted''
triangular lattice (Fig.~7b)
or 
the 
QP Penrose lattice itself
are 
good candidates for
the optimal enhancement of the critical current 
in the regions 
where 
the function 
$J_{c}(\Phi)$
have minima for a periodic lattice. 
(It should be noted here
that the curves shown, 
e.g., 
in Fig.~7h,
for the triangular and the Penrose lattices
are calculated for the 
{\it same }
cell, 
although the effective area of the Penrose lattice
is 
{\it smaller }
than that of the triangular lattice
with the same number of pinning sites. 
This discrepancy is taken into account by 
the ``filling factor'' introduced by Eq.~(\ref{eta}).
It should be also recalled, 
when comparing 
the function 
$J_{c}(\Phi)$
for the case of the triangular and the Penrose lattices, 
that 
the main maximum of the curve for the Penrose lattice
(see Fig.~11) is the second sharp peak 
at
$\Phi = \Phi_{1}$.)

In this respect, 
it is interesting to compare the above results for 
the 
``quasiperiodic distortion''
of the triangular lattice 
with 
its random distortion. 
For this purpose, 
we introduce 
a random angle 
$\alpha_{\rm ran}$: 
$0 < \alpha_{\rm ran} < 2\pi$,
and 
a random radius of the displacement 
$d_{\rm ran}$: 
$0 < r_{\rm ran} < r_{\rm ran}^{\rm max}$,
where 
$r_{\rm ran}^{\rm max}$
is the maximal displacement radius, 
which is a measure of noise
measured in units of 
$a/2$, 
where 
$a$ 
is the (triangular) lattice constant.

In Fig.~13, 
randomly distorted triangular lattices are shown 
for 
$r_{\rm ran}^{\rm max} = 0.2 (a/2)$ (Fig.~13a), 
$0.3 (a/2)$ (Fig.~13b), 
$0.4 (a/2)$ (Fig.~13c), 
and 
for 
$a/2$ (Fig.~13d). 
For comparison, 
the triangular lattice is also shown in Figs.~14a to d. 
The corresponding functions 
$J_{c}(\Phi)$
are presented,  
respectively, 
in Figs.~13e to h. 
At low levels of noise (e.g., Fig.~13e, f) 
the valleys (minima) start to fill 
due to the disappearance of the long-range order, 
similarly to the


\begin{figure}[btp]
\begin{center}
\end{center}
\caption{ 
(This figure is available in ``png'' format: ``Penrose Fig 13.png''; color)
Left column: 
Randomly distorted triangular lattices for 
$r_{ran}^{max} = 0.2 (a/2)$ (a), 
$0.3 (a/2)$ (b),
$0.4 (a/2)$ (c),
and 
for
$a/2$ (d). 
For comparison, 
the triangular lattice is also shown. 
Right column: 
The functions 
$J_{c}(\Phi)$
corresponding
to the distributions of pinning sites 
shown in (a) to (d), 
are presented,  
respectively, 
in (e) to (h). 
At low levels of noise (e), (f), 
the valleys (minima) start to fill 
due to disappearance of the long-range order, 
similarly to the case of the Penrose lattice, 
although accompanied with a weaker enhancement of
$J_{c}$
than for the case of the 
``quasiperiodic distortion'' (Fig.~7f).
For higher levels of noise, 
the main peak degrades 
without any essential enhancement of 
$J_{c}$
in the neighborhood (g), (h). 
}
\end{figure}



\begin{figure}[btp]
\begin{center}
\vspace*{-7.0cm}
\hspace*{-1.0cm}
\includegraphics*[width=10.5cm]{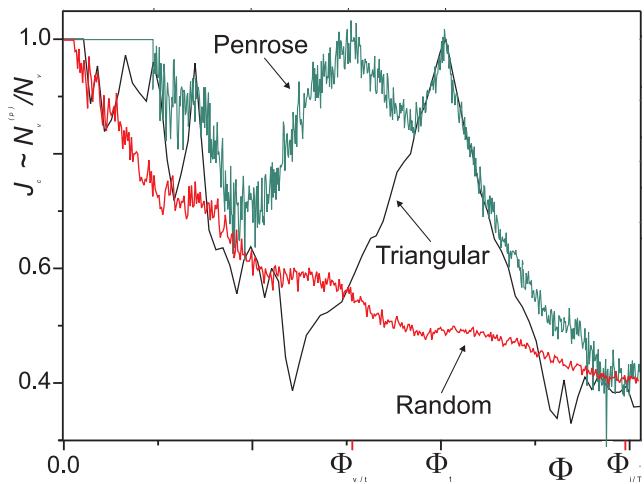}
\end{center}
\vspace{-1.0cm}
\caption{
(Color)
The critical current 
$J_{c}(\Phi)$
for a 301-sites Penrose-lattice (dark green solid line),
(recalculated for flux {\it only} on the Penrose area, $A_{P}$),
triangular (black solid line)
and random (red solid line) 
pinning arrays.
The Penrose lattice provides a remarkable enhancement 
of $J_{c}(\Phi)$
over a very wide range of values of
$\Phi$ because it contains many periods in it.
}
\end{figure}


\noindent 
case of the Penrose lattice, 
although accompanied with a weaker enhancement of 
$J_{c}$ than for the case of the 
``quasiperiodic distortion'' (Fig.~7f). 
For higher levels of noise, 
the main peak degrades 
without any essential enhancement of 
$J_{c}$ 
in the neighborhood (Figs.~13g, h).

For comparison, 
we also show 
in Fig.~14 
the 
$J_{c}(\Phi)$ 
for a Penrose-lattice 
(calculated for the sample with $N_{p}=301$, only for the area of the Penrose lattice $A_{P}$, 
see Eq.~(\ref{eta})). 
Notice that the QP lattice leads to a very broad and potentially useful 
{\it enhancement} 
of the critical current $J_{c}(\Phi)$, 
even compared to the triangular or random APS. 
The remarkably broad maximum in 
$J_{c}(\Phi)$ 
is due to the fact that the Penrose lattice has 
{\it many} 
(infinite, in the thermodynamic limit) 
periodicities built in it \cite{bookqc}. 
In principle, 
each one of these periods provides a peak in $J_{c}(\Phi)$. 
In practice, like in quasicrystalline 
difraction patterns, only few peaks sre strong. 
This is also consistent with our study. 
Furthermore, the pinning parameters can be 
adjusted by using as pinning centers either antidots 
(microholes of different radii ``drilled'' in the film 
\cite{vvmdotprl,rwdot}, or blind antidots \cite{vvmbdot} of
different depths and radii. Thus, our results could be 
observed experimentally.

\section{Conclusions}

The critical depinning current $J_{c}$, 
as a function of the applied magnetic flux $\Phi$, 
has been studied in QP pinning arrays, 
from one-dimensional chains to 
two-dimensional arrays of pinning sites set in the nodes of 
quasiperiodic lattices
including 
a 2D-quasiperiodic triangular lattice and 
a five-fold Penrose lattice.

In a 1D quasiperiodic chain of pinning sites,
positions of the peaks of the function 
$J_{c}(\Phi)$
are governed by ``harmonics'' of long and short periods 
of the quasiperiodic chain. 
Independently of the length of the chain, 
the peaks form a set of quasiperiodic sequence
including a Fibonacci sequence as a basic subset. 
Analyzing the evolution of the peaks, 
when a continuous transition is performed from a periodic 
to a quasiperiodic lattice of the pinning sites, 
we found that 
the peaks related to the Fibonacci sequence are most pronounced when the ratio of lengths 
of the long and the short periods is the golden mean. 
A comparison of the sets of peaks 
for different chains 
shows that 
the functions
$J_{c}(\Phi)$
for the 1D quasiperiodic chain 
is 
self-similar. 
In the $k$-space,
the self-similarity effect is displayed 
in the Fourier-transform of the distribution function
of the system of vortices pinned on a 1D quasiperiodic 
array of pinning centers. 
The evolution of quasiperiodic peaks
when gradually changing the ``quasiperiodicity'' parameter 
$\gamma$
(i.e., ratio of the lengths of short to long elements of a quasiperiodic chain)
has revealed a continuous transition from a periodic chain --
through the set of quasiperiodic states --
to another periodic chain with a longer period. 
This phenomena has been studied both in real space 
and in reciprocal $k$-space. 

In 2D quasiperiodic pinning arrays (e.g., Penrose lattice), the pinning of vortices is 
related to matching conditions between a triangular vortex lattice and the quasiperiodic 
lattice of the pinning centers. 
Although more complicated than in 1D pinning chains, the specific behavior of $J_{c}(\Phi)$
is determined by the presence of two different kinds of elements -- 
thick and thin rhombuses --
forming the quasiperiodic lattice. 
Based on these considerations, 
the positions of the main maxima of $J_{c}(\Phi)$ for Penrose lattice are predicted. 

In particular, 
for the first matching field 
each pinning site is occupied by a vortex. 
The corresponding maximum 
of the function 
$J_{c}(\Phi)$ 
is broad since it involves at least three kinds of local 
matching effects of the flux lattice, 
with the rhombus side and 
with short diagonals of thick and thin rhombuses. 

Another Penrose-lattice 
matching field 
is related with
local matching effects 
which involve 
the intervortex distance of the vortex flux lattice, 
the rhombus side 
and 
the short diagonal of thick rhombus. 
For this field, 
all the pinning sites are occupied, 
which are 
situated in the vertices of thick rhombuses 
and only 
three out of four 
in the vertices of thin rhombuses.
The number of unoccupied pinning sites is governed by the number of thin rhombuses. 
Some of the thin rhombuses are single 
(i.e., separated from other thin rhombuses by thick ones), 
while some of them are double (i.e., have common sides with each other). 
Therefore, the number of vacancies is the number of single thin rhombuses 
plus one half of the number of ``double'' thin rhombuses. 
One more important feature of the function 
$J_{c}(\Phi)$ occurs for higher vortex densities, 
when a single interstitial vortex enters each thick rhombus.  

Numerical simulations performed for various sample sizes have revealed 
a good agreement with our predictions. 

The revealed features can be more or less pronounced depending on 
specific relations between 
the vortex-vortex interaction constant and 
the strength of the pinning sites, 
as well as on the distance between pinning sites
and their radius. 
While 
the vortex-vortex interaction constant is a material parameter, 
all others 
can be adjustable parameters in experiments with
artificially created quasiperiodic pinning arrays. 
This can be reached by using, for instance, antidots
(i.e., microholes ``drilled'' in a superconductor film)
or blind antidots of different depths and radii 
as pinning centers. 
Our calculations provide the necessary
relations between these parameters
for possible experimental realizations. 

A continuous deformation of the Penrose lattice to a periodic triangular lattice 
(i) shows that the above revealed features are hallmarks of quasiperiodic pinning arrays; 
(ii) provides us with a tool for the controlled change of the magnitude, 
sharpness and the position of the peaks of $J_{c}(\Phi)$ that is important for possible 
applications. 
In particular,
our analysis shows that the quasiperiodic lattice 
provides an unusually broad critical current $J_{c}(\Phi)$, 
that could be useful for practical applications 
demanding high 
$J_{c}$'s 
over a wide range of fields.

\vspace*{1.0cm}
\vspace*{1.0cm}
\vspace*{1.0cm}

\section*{ACKNOWLEDGMENT}

This work was supported in part by the National Security Agency 
(NSA) and Advanced Research and Development Activity (ARDA) under
Air Force Office of Scientific Research (AFOSR) contract number
F49620-02-1-0334; and also supported by the US National Science
Foundation grant No.~EIA-0130383, and RIKEN's President's funds.

\section*{APPENDIX A: Onset of vortex motion for currents higher 
then the critical current: $J > J_{c}$}

Here we present 
vortex flow patterns for currents exceeding the critical value, $J > J_{c}$. 
In Fig.~15, 
the vortex flow patterns are shown calculated for a Penrose sample with $N_{p} = 141$  
(pinning sites shown by red circles). 
The ground-state vortex configuration is shown in Fig.~15a (vortices shown by green dots) 
for $\Phi \approx \Phi_1$ and when no driving force is applied. 
This vortex configuration is similar to those shown in Figs.~9c and 10c, when the number of 
vortices (within the sample area) is equal to the number of pinning sites, and all the 
vortices are pinned. 
When an increasing driving force $f_{d} \sim J$ is applied, the vortices do not move 
until $f_{d}$ reaches some threshold value, when vortices depin. 
The current which corresponds to the driving force depinning the vortices, 
is then defined as the critical current $J_{c}$ (dynamical criterion). 
A comparison of $J_{c}$'s calculated using this criterion and the static criterion, 
is shown above in Fig.~11b. 
When unpinned, the vortices move along complicated trajectories, or ``channels'' 
created by pinning arrays and interacting with other vortices. 
Figs.~15b to 15d show the onset of the flux motion for $J > J_{c}$, following the 
traces of moving vortices over distances about 
$0.5\lambda$ (b), $1\lambda$ (c), and $2\lambda$ (d). 
On the subsequent consecutive snapshots, vortices are shown by a sequence of consecutive 
open blue circles (and by blue solid circles for the last snapshot on each panel). 
These show dynamical configurations of the vortex lattice in motion. 
Note the appearance of local ``rivers'' of vortices moving along the channels between 
neighboring pinning sites.


\begin{figure}[btp]
\begin{center}
\end{center}
\caption{ 
(This figure is available in ``png'' format: ``Penrose Fig 15.png''; color)
Vortex flow patterns for $J > J_{c}$, 
calculated for a Penrose sample with $N_{p} = 141$ (pinning sites shown by red circles). 
(a) Ground-state vortex configuration (vortices shown by green dots) when $\Phi \approx \Phi_1$, and no driving force is applied. 
(b-d) The onset of the flux motion for $J > J_{c}$, following the traces of moving vortices over distances about 
$0.5\lambda$ (b), $1\lambda$ (c), and $2\lambda$ (d). 
On the subsequent consecutive snapshots, vortex trajectories are shown by 
black dotted lines 
(the blue solid circles show the last snapshot on each panel). 
}
\end{figure}



\begin{references}

\bibitem{vvmdotprl}
M.~Baert, V.V.~Metlushko, R.~Jonckheere, V.V.~Moshchalkov, and Y.~Bruynseraede, Phys. Rev. Lett. {\bf 74}, 3269 (1995).
\bibitem{vvmdotprb}
V.V.~Moshchalkov, M.~Baert, V.V.~Metlushko, E.~Rosseel, M.J.~Van~Bael, K.~Temst, R.~Jonckheere, and Y.~Bruynseraede, 
Phys. Rev. B {\bf 54}, 7385 (1996).

\bibitem{ulm01}
J.~Eisenmenger, P.~Leiderer, M.~Wallenhorst, and H.~D\"{o}tsch, 
Phys. Rev. B {\bf 64}, 104503 (2001). 
\bibitem{ulm02}
J.~Eisenmenger, Z.-P.~Li, W.A.A.~Macedo, and I.K.~Schuller, 
Phys. Rev. Lett. {\bf 94}, 057203 (2005). 

\bibitem{fnsc2003} 
J.E. Villegas, S. Savel'ev, F. Nori, E.M. Gonzalez, J.V. Anguita, R. Garc{\'i}a, and J.L. Vicent, 
Science {\bf 302}, 1188 (2003);
J.E.~Villegas, E.M.~Gonzalez, M.I.~Montero, I.K.~Schuller, J.L.~Vicent, 
Phys. Rev. B {\bf 68}, 224504 (2003); 
M.I.~Montero, J.J.~Akerman, A.~Varilci, I.K.~Schuller, 
Europhys. Lett. {\bf 63}, 118 (2003). 

\bibitem{rwdot}
A.M.~Castellanos, R.~W\"{o}rdenweber, G.~Ockenfuss, A.v.d.~Hart, and K.~Keck, Appl. Phys. Lett. {\bf 71}, 962 (1997); 
R. W\"{o}rdenweber, P. Dymashevski, and V.R. Misko, Phys. Rev. B {\bf 69}, 184504 (2004).

\bibitem{vvmbdot}
L.~Van~Look, B.Y.~Zhu, R.~Jonckheere, B.R.~Zhao, Z.X.~Zhao, and V.V.~Moshchalkov, 
Phys. Rev. B {\bf 66}, 214511 (2002).
%
%
\bibitem{vvmfddot}
A.V.~Silhanek, S.~Raedts, M.~Lange, and V.V.~Moshchalkov, 
Phys. Rev. B {\bf 67}, 064502 (2003).

\bibitem{penrose}
R.~Penrose, Bull. Inst. Math. Appl. {\bf 10}, 226 (1974);
R.~Penrose, Math. Intelligencer {\bf 2} (1), 32 (1979). 
\bibitem{debruijn}
N.G.~de~Bruijn, Koninklijke Ned. Akad. Weten. Proc., 
Ser. A {\bf 84}, 39 (1981); {\bf 84}, 53 (1981).


\bibitem{bookqc}
{\it Quasicrystals}, Ed. J.-B.~Suck, M.~Schreiber, P.~H\"{a}ussler 
(Springer, Berlin, 2002). 

\bibitem{zia}
R.K.P.~Zia and  W.J.~Dallas, 
J. Phys. A: Math. Gen. {\bf 18}, L341 (1985). 


\bibitem{kohmoto}
M.~Kohmoto, B.~Sutherland, and C.~Tang, 
Phys. Rev. B {\bf 35}, 1020 (1987).

\bibitem{fnphon1d}
F.~Nori and J.P.~Rodriguez, 
Phys. Rev. B {\bf 34}, 2207 (1986).


\bibitem{fnniu86}
Q.~Niu and F.~Nori, 
Phys. Rev. Lett. {\bf 57}, 2057 (1986).

\bibitem{behrooz}
A.~Behrooz, M.J.~Burns, H.~Deckman, D.~Levine, B.~Whitehead, and P.M.~Chaikin,
Phys. Rev. Lett. {\bf 57}, 368 (1986); 
Y.Y.~Wang R.~Steinmann, J.~Chaussy, R.~Rammal, and B.~Pannetier, 
Jpn. J. Appl. Phys. {\bf 26}, 1415 (1987); 
K.N.~Springer and D.J.~Van~Harlingen, 
Phys. Rev. B {\bf 36}, 7273 (1987). 

\bibitem{fnniu87}
F.~Nori, Q.~Niu, E.~Fradkin, and S.-J.~Chang,
Phys. Rev. B {\bf 36}, 8338 (1987).
\bibitem{fnniu88}
F.~Nori and Q.~Niu, 
Phys. Rev. B {\bf 37}, 2364 (1988).
\bibitem{niufn89}
Q.~Niu and F.~Nori,
Phys. Rev. B {\bf 39}, 2134 (1989).
\bibitem{linfn02}
Y.-L.~Lin and F.~Nori,
Phys. Rev. B {\bf 65}, 214504 (2002).


\bibitem{zhu}
S.-N.~Zhu, Y.-Y.~Zhu, and N.-B.~Ming, 
Science {\bf 278}, 843 (1997). 
\bibitem{kivshar}
F.~Dom\'{\i}nguez-Adame, A.~S\'{a}nchez, and Y.S.~Kivshar, 
Phys. Rev. E {\bf 52}, R2183 (1995).
\bibitem{torres}
M.~Torres, J.P.~Adrados, J.L.~Aragon, P.~Cobo, and S.~Tehuacanero, 
Phys. Rev. Lett. {\bf 90}, 114501 (2003).

\bibitem{md01}
F.~Nori, Science {\bf 278}, 1373 (1996);
C.~Reichhardt, C.J.~Olson, J.~Groth, S.~Field, and F.~Nori, Phys. Rev. B {\bf 52}, 10~441 (1995); 
B {\bf 53}, R8898 (1996);
B {\bf 54}, 16~108 (1996);
B {\bf 56}, 14~196 (1997);
\bibitem{md0157}
C.~Reichhardt, C.J.~Olson, and F.~Nori, Phys. Rev. B {\bf 57}, 7937 (1998).
\bibitem{md02}
C.~Reichhardt, C.J.~Olson, and F.~Nori, Phys. Rev. Lett. {\bf 78}, 2648 (1997); 
B {\bf 58}, 6534 (1998). 
\bibitem{md03Z}
B.Y.~Zhu, F.~Marchesoni, V.V.~Moshchalkov, and F.~Nori, Phys. Rev. B {\bf 68}, 014514 (2003); 
Physica C {\bf 388-389}, 665 (2003);
Physica C {\bf 404}, 260 (2004); 
B.Y.~Zhu, L. Van Look, F.~Marchesoni, V.V. Moshchalkov, and F. Nori, 
Physica E {\bf 18}, 322 (2003); 
B.Y.~Zhu, F.~Marchesoni, and F.~Nori, Phys. Rev. Lett. {\bf 92}, 180602 (2004); 
Physica E {\bf 18}, 318 (2003); 
F.~Marchesoni, B.Y.~Zhu, and F. Nori, Physica A {\bf 325}, 78 (2003). 


\bibitem{kolarfn90}
M.~Kol\'{a}\c{r} and F.~Nori, 
Phys. Rev. B {\bf 42}, 1062 (1990).

\bibitem{tonomura-vvm}
K.~Harada, O.~Kamimura, H.~Kasai, T.~Matsuda, A.~Tonomura, and V.V.~Moshchalkov, 
Science {\bf 274}, 1167 (1996).

%
%

%
%

\bibitem{togawa}
Y.~Togawa et al. 
Phys. Rev. Lett. {\bf 95}, in press (2005).

\bibitem{brandt}
E.-H.~Brandt
J. Low Temp. Phys. {\bf 26}, 709, 735 (1977); 
Phys. Rev. B {\bf 34}, 6514 (1986); 
Rep. Prog. Phys. {\bf 58}, 1465 (1995). 

\bibitem{vinokur} 
G.~Blatter, M.V.~Feigel'man, V.B.~Geshkenbein, A.I.~Larkin, and V.M.~Vinokur, 
Rev. Mod. Phys. {\bf 66}, 1125 (1994). 

\bibitem{clem} 
J.R.~Clem, M.W.~Coffey, 
Phys. Rev. B {\bf 46}, 14622 (1992). 

\end{references}
\end{document}